\documentclass[ prd,twocolumn]{revtex4}
\usepackage{xcolor}
\usepackage[utf8]{inputenc}
\usepackage{graphicx}
\usepackage{empheq}
\usepackage{braket}
\usepackage{float}
\usepackage{amssymb}
\usepackage{amsmath}
\usepackage{mhchem}
\usepackage{siunitx}
\usepackage[titletoc]{appendix}

\usepackage{hyperref}
\hypersetup{colorlinks=true, linkcolor=red, citecolor=blue, urlcolor=purple}

\begin{document}

\title {Chiriality in a three-band superconducting prism in ZFC and FC processes}

\author{C. A. Aguirre$^{\dagger}$\href{https://orcid.org/0000-0001-8064-6351}{\includegraphics[scale=0.05]{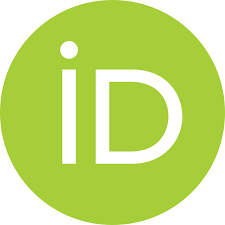}}}
\affiliation{Departamento de Física, Universidade Federal de Mato-Grosso, Cuiabá, Brasil.\\
Condensed Matter Physics Group, Instituto de Fisica, Universidade Federal do Rio Grande do Sul, $91501-970$
Porto Alegre, RS, Brazil.}

\author{Julián Faúndez\href{https://orcid.org/0000-0002-6909-0417}{\includegraphics[scale=0.05]{Figures/orcid.png}} and S. G. Magalhães\href{https://orcid.org/0000-0002-6874-7579}{\includegraphics[scale=0.05]{Figures/orcid.png}} }

\affiliation{Condensed Matter Physics Group, Instituto de Fisica, Universidade Federal do Rio Grande do Sul, $91501-970$
Porto Alegre, RS, Brazil}

\author{J. Barba-Ortega\href{https://orcid.org/0000-0003-3415-1811}{\includegraphics[scale=0.05]{Figures/orcid.png}}}
\affiliation{Departamento de Física, Universidad Nacional de Colombia, Bogotá, Colombia\\ Foundation of Researchers in Science and Technology of Materials, Bucaramanga, Colombia}

\date{\today}
\begin{abstract} 
In the present work, we will study the superconducting properties of a mesoscopic three-dimensional prism with square transversal section under an external magnetic field in both Zero-Field Cooling (ZFC) and Field-Cooling (FC) process. The studied sample is a three-band system (with three condensates $\psi_{1}$, $\psi_{2}$, $\psi_{3}$), in which we take a Josephson type inter-band coupling simulated via $\gamma$ parameter. We analyze the magnetization, superconducting electronic density and free Gibbs energy considering both processes. As results, we found an interesting tunneling of vortex and anti-vortex between the bands, and vortex cluster configuration in the sample. This behaviour is due to the non-monotonic interaction between the bands.
\end{abstract}
\maketitle
\section{Introduction}
During the last decades the study of highly correlated systems (HCS) has taken on great relevance within the scientific community, due to the discovery of high critical temperature superconductors (HTSc) \cite{HTS1,HTS2,HTS3}, such as cuprates (copper oxide) \cite{CS1,CS2,CS3}, compounds of heavy fermions \cite{FP1,FP2}, iron-based superconductors (iron pnictides) \cite{ib1,ib2}, organic superconductors \cite{OS1,OS2}, super/rotated graphene lattices \cite{GR1,GR2}, until the most recent discovery of superconductivity in nickel (nickel oxides) \cite{Li,zeng}, this in addition to multiple measurements of new vortex states in un-conventional superconductors; these measurements have been made using different experimental techniques, such as \textit{ARPES} \cite{Arpes}, Bitter decoration \cite{Bitter} and Squids \cite{Squid}. Through which they have shown that in multi-band systems, the behavior of the vortices is unconventional and they tend to generate conglomerates or clusters \cite{Cluster}, this behavior being the result of the interaction between the crystal lattice (optical phonons) and the Cooper pairs coupled in the superconducting system, in each of the bands \cite{Cluster2}.
This has opened a new era in the study of superconductors in multiple technological applications, ranging from measurements of mild magnetic fields (Squids) \cite{Barba2010,Barba2011} to important applications in medicine, spinor \cite{Medi1}, speed and data processing \cite{Medi2}, technological applications in memory and new and improved processing speeds of large databases \cite{Medi3}, due to the intrinsic properties of said non-conventional superconductors. With this, there are essentially several approaches for the study of the superconducting phase (superconducting gap), which are grouped in microscopic terms, based on the Bardeen-Cooper-Schieffer (BCS) theory and respective expansions as Migdal-Eliasberg or Bogoliubov-DeGenns \cite{Migdal, Bogoliubov}, Ab-initio studies based on its atomic (or molecular) structure and band structure \cite{Abinitio1} and finally through the phenomenological study, mediated by the time dependent Ginzburg-Landau (TDGL) theory command parameter\cite{Multi}. Recent discoveries in new unconventional superconducting materials have generated a renewed interest in new interesting topological phases\cite{PhysicaB2021}, such as multi-band effects (multi-condensed) \cite{JLTP2021,a5}, mesoscopic superconductivity \cite{Cabral2004,Clecio2001}, fractional vorticity \cite{Egora}, kinematic vortices \cite{Sardela2000} and vortex clusters due to repulsive short-range and attractive long-range interaction \cite{Egor2,Egor3}.
Thus, the study of multi-band systems has become essential to capture the essential physics in certain materials of interest such as $MgB_{2}$ \cite{Yuriy,Misko2003}, which presents multiple gaps in the superconducting excitation spectrum \cite{Canfield,Szabo,Iavarone}, also in $Sr_{2}RuO_{4}$ which in its pure state is one of the best candidates to constitute three-band superconducting order parameters \cite{Maeno,Baskaran,Agterberg}. In this way, each physical system must be studied specifically depending on the number of existing and interacting gaps observed, it is also essential to describe the nature of those interactions between the superconducting order parameters that they describe in material and more importantly, establish the type of coupling, which actually mediates the interaction between the order parameters. With this, we propose the use of a Josephson-type coupling between the order parameters in a three-dimensional superconducting mesoscopic prism composed of three interacting order parameters with symmetry breaking $Z_{[{0,2\pi}]}$ \cite{Cluster2}. 
This article is organized as follows: the theoretical formalism is presented in section \ref{Section1}. In section \ref{Section22} we present the main results for the proposed Josephson-type coupling and whose coupling force will be mediated by the parameter $\gamma$ in the ZFC and FC processes. We show the superconducting order parameter $|\psi_{1}|,|\psi_{2}|,|\psi_{3}|$, magnetization, Gibbs free energy as a function of temperature and external, the phase of the order parameter show an evidence of the tunneling of vortex and anti-vortex between the bands. Finally, in section \ref{Section3} we detail the main results. 
\section{Theoretical Formalism}\label{Section1}
\begin{figure}[htbp!]
\centering
\includegraphics[scale=0.45]{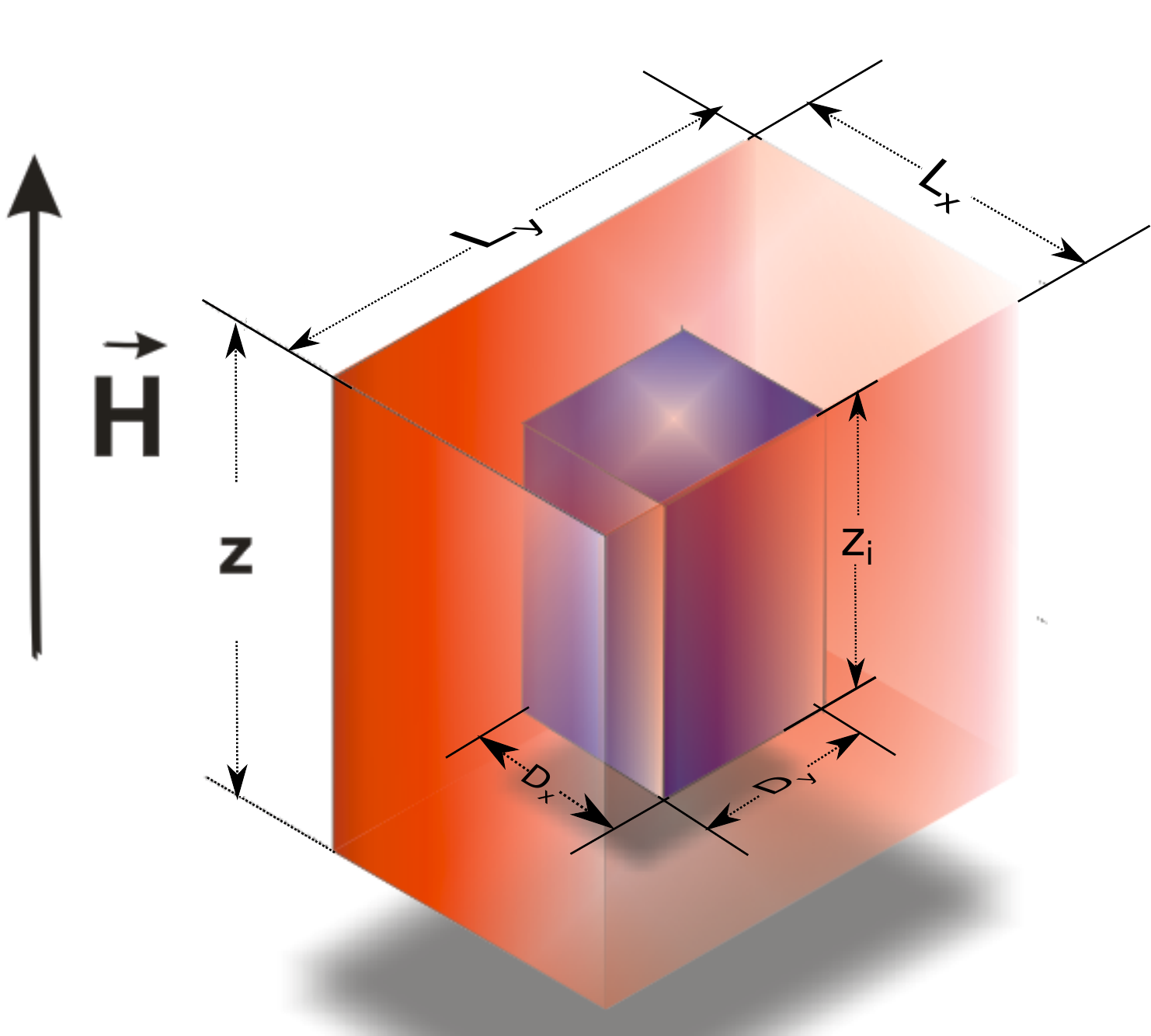}
\caption{Layout of the superconducting studied three-band prism. The size of the computational mesh is $L_x=L_y=50\xi_{1}$, $Z=25\xi_{1}$, and for the internal superconducting box is $D_x=D_y=25\xi_{10}$, $z_l=12.5\xi_{10}$.} 
\label{Sample}
\end{figure}

In this work, we will study a superconducting nano-prism, through the functional of a three-band system. We will consider the interaction between the three bands, in Josephson type coupling. Thus, the Gibbs energy density for the the superconducting order parameter complex pseudo-function $\psi_{i}=|\psi_{i}|e^{i\theta_{i}}$ ($\theta_{i}$ its phase), and magnetic potential $\mathbf{A}$ is \cite{TDGL2B,TDGL2B2,TDGL2B3}:
\begin{eqnarray}
\mathcal{G} = \int dV ( \sum_{i}^{3} \mathcal{F}(\psi_{i},\textbf{A})
+\frac{1}{2 \mu_0} |\nabla \times \mathbf{A}|^2+\Theta(\psi_{i}))\label{Gibbs1}
\end{eqnarray}
where:
{\small
\begin{equation}
\mathcal{F}(\psi_{i},\textbf{A})= \alpha_i |\psi_i|^2 +\frac{\beta_i}{2} |\psi_i|^4+\frac{\zeta_{i}}{2 m_i} |(i \hbar \nabla+2 e \mathbf{A}) \psi_i |^2
\label{Function}
\end{equation}}
and
\begin{eqnarray}
\Theta(\psi_{1},\psi_{2},\psi_{3})=\gamma (\psi_1^*\psi_2\psi_3+\psi_1 \psi_2^*\psi_3^*) +\label{josep} \\  \gamma (\psi_1\psi_2^*\psi_3+\psi_1^{*} \psi_2\psi_3^*)\nonumber + \gamma (\psi_1\psi_2\psi_3^*+\psi_1^* \psi_2^*\psi_3)
\end{eqnarray}
 $\alpha_i=\alpha_{i0}(1-T/T_{ci})$ and $\beta_i $ are two phenomenological parameters, $i=1,2,3$ in the equations \ref{Gibbs1} and \ref{Function}. The equation \ref{josep} describes our Josephson coupling proposal. The inclusion of the parameter $\zeta_{i}$ in the equation \ref{Function} indicating  the height of the super-currents in each band. In the London gauge $\nabla \cdot \mathbf{A}=0$, The general form of time dependent Ginzburg-Landau equations for a three-band (3B-TDGL) system in dimensionless units are given by \cite{PhysicaB2021,Egor2,Egor3}:
\begin{align}
\frac{\partial \psi_{1}}{\partial t}=(1-T-|\psi_1|^2) \psi_1-|i \nabla+\mathbf{A}|^2 \psi_1 +\hat{\gamma}_{23}
\label{GL2B1}
\end{align}
\begin{eqnarray}
\frac{\partial \psi_{2}}{\partial t}=(1-\frac{T}{T_{r2}}-|\psi_2|^2)\psi_2- 
\frac{m_{r2}}{\alpha_{r3}} |i \nabla+\mathbf{A}|^2 \psi_2 +\hat{\gamma}_{13}
\label{GL2B2}
\end{eqnarray}
\begin{eqnarray}
\frac{\partial \psi_{3}}{\partial t}=(1-\frac{T}{T_{r3}}-|\psi_3|^2) \psi_3-\frac{m_{r3}}{\alpha_{r2}}|i \nabla+\mathbf{A}|^2\psi_3 + \hat{\gamma}_{12} 
\label{GL2B3}
\end{eqnarray}
where:
\begin{eqnarray}
  \label{GL4}
 \hat{\gamma}_{23}=\gamma|\psi_{2}||\psi_{3}|\epsilon^{i\theta_{1}}\Big[\cos(\mu_{23})+3i\sin(\mu_{23})\Big]
  \end{eqnarray}
\begin{eqnarray}
 \hat{\gamma}_{13}=\sqrt\frac{\alpha_{2}}{\beta_{2}}\frac{\alpha_{r2}^2}{\beta_{r2}}\gamma|\psi_{1}||\psi_{3}|\epsilon^{i\theta_{2}}\Big[\cos(\mu_{13})+3i\sin(\mu_{13})\Big] 
  \end{eqnarray}
\begin{eqnarray}
 \hat{\gamma}_{12} =\sqrt\frac{\alpha_{3}}{\beta_{3}}\frac{\alpha_{r3}^2}{\beta_{r3}} \gamma|\psi_{1}||\psi_{2}|\epsilon^{i\theta_{3}}\Big[\cos(\mu_{12})+3i\sin(\mu_{12})\Big]
 \end{eqnarray}
and
\begin{eqnarray}
 \mu_{23}=\theta_{2}+\theta_{3}-\theta_{1}\\
 \mu_{13}=\theta_{1}+\theta_{3}-\theta_{2}\\
  \mu_{12}=\theta_{1}+\theta_{2}-\theta_{3}\label{mu}
 \end{eqnarray}
 and the equation for the vector potential is:
\begin{align}
\frac{\partial \mathbf{A}}{\partial t}=\zeta_{1}\Re \left[\psi_1(i\nabla -\mathbf{A}) \psi_1^*  \right]+ \nonumber \\
\zeta_{2} \Re \left[\frac{\beta_{r2}}{\alpha_{r3}}\psi_2(i\nabla -\mathbf{A}) \psi_2^*\right]+ \nonumber \\
\zeta_{3} \Re \left[ \frac{\beta_{r3}}{\alpha_{r2}}\psi_3 (i \nabla -\mathbf{A}) \psi_3^*\right]+ \kappa^2 \triangle \mathbf{A}
\label{GL2A}
\end{align}
$\hat{\gamma}_{ij}$ from the equation \ref{GL2B1} until the equation \ref{mu},  represents the Josephson coupling between the $i$ and $j$ band. The boundary conditions ${\bf n}\cdot(i\mbox{\boldmath$\nabla$}+{\bf A})\psi_{i}=0$, $i=1,2,3$ with ${\bf n}$ a surface normal outer vector. Also, we defined $T_{r2}=T_{3}/T_{2}=1.0$, $T_{r3}=T_{r2}^{-1}$; 
$\alpha_{r2}=\alpha_{30}/\alpha_{20}=0.7$, $\alpha_{r3}=\alpha_{r2}^{-1}$, $m_{r2}=m_{2}/m_{3}=0.5$, $m_{r3}=m_{r2}^{-1}$, $\beta_{r2}=\beta_{3}/\beta_{2}=0.7$,$\beta_{r3}=\beta_{r2}^{-1}$.
We express the temperature $T$ in units of the critical temperature $T_{c1}$, length in units of the coherence length $\xi_{10}=\hbar/\sqrt{-2 m_1 \alpha_{10}}$, the order parameters in units of $\psi_{i0}=\sqrt{-\alpha_{i0}/\beta_i}$ and the Ginzburg-Landau parameter $\kappa=1.0$. We choose the zero-scalar potential gauge at all times and use the link variables method for to solve the 3B-TDGL equations \cite{Sardela2000, TDGL2B,TDGL2B2,TDGL2B3}(and references therein). Finally, for convergence rule for time:
\begin{eqnarray}
\Delta t \leq min \bigg\{\frac{a^2}{4}, \frac{\beta a^2}{4\kappa^2}\bigg\}; \qquad a^2 = \frac{2}{\frac{1}{\delta x^2} + \frac{1}{\delta y^2}+\frac{1}{\delta z^2}}
\end{eqnarray}
Time is in units of the Ginzburg-Landau characteristic time $t_{GL}=\pi\hbar /8k_BT_{c1}$, and the vector potential $\mathbf{A}$ is scaled by $H_{c2}\xi_{10}$, where $H_{c2}$ is the bulk upper critical field. Gibbs free energy $G$ in $G_0=H_c^{2}\xi_{10}/4\pi \epsilon\xi_{10}^{2}$ units. We present the layout of the superconducting sample in Figure \ref{Sample}. 
With this, it's important to establish the possible chilarities that the studied coupling can fulfill. Since the parameter $\hat{\gamma}$ defines the possibility of tunneling between the Cooper pairs present in each band, the two possible options $\hat{\gamma}_{ij}>0$ and $\hat{\gamma}_{ij}<0$ can be linked in different ways in the Eq. \ref{GL2B1} to Eq. \ref{GL2B3}. Also by means of the parameter $\zeta_{i}>0$ and $\zeta_{i}<0$ in the Eq. \ref{GL2A}, with this, we present the six different possibilities of combination of these two parameters in the
Table \ref{Table1}. We use $|\gamma|=0.01$, and considered the band $1$ as a dominant band, therefore $|\zeta_1|=1.0$, and $|\zeta_2|=|\zeta_3|=0.01$, for all studied cases (for more details of these calculus see the appendix \ref{apendice1}).
\begin{center}
\begin{table}
\begin{tabular}{ |c|  |c|  |c|  |c|  |c|  |c|  |c| }
\hline
 Case                        &  I  &  II & III & IV  &  V  &  VI  \\ 
\hline
  $\hat{\gamma}_{23}$        & $+$ & $+$ & $+$ & $+$ & $+$ & $+$  \\ 
\hline\hline
 $\hat{\gamma}_{13}$         & $+$ & $-$ & $+$ & $+$ & $-$ & $+$  \\ 
\hline\hline
 $\hat{\gamma}_{12}$         & $+$ & $-$ & $+$ & $+$ & $+$ & $+$  \\ 
\hline\hline
 $\zeta_1$                   & $+$ & $-$ & $+$ & $-$ & $-$ & $+$ \\ 
\hline\hline
$\zeta_2$                    & $+$ & $-$ & $+$ & $+$ & $+$ & $-$ \\ 
\hline\hline
 $\zeta_3$                   & $+$ & $-$ & $+$ & $+$ & $-$ & $-$ \\ 
\hline\hline
\end{tabular}
\caption{Studied combinations of $\hat{\gamma}_{ij}$ and $\zeta_i$ values, for chirality in the order parameter and super-current. $|\gamma|=0.01$, $|\zeta_1|=1.0$ and $|\zeta_2|=|\zeta_3|=0.01$.}
\label{Table1}
\end{table}
\end{center}
\section{Results and discussion}\label{Section22}
In the following sections, we will present the order parameter and its phase in each band, magnetization and Gibbs energy, for all cases presented in the table \ref{Table1}. Now, the study will be carried out via the ZFC and FC processes. For the ZFC process, we choose $H=0.9$ and $H=1.1$. For the FC process all our simulations are performed taking $T=0$.
\subsection{Zero-Field-Cooling process at $H$ constant.}
\begin{figure}[htbp!]
\centering
\includegraphics[scale=0.3]{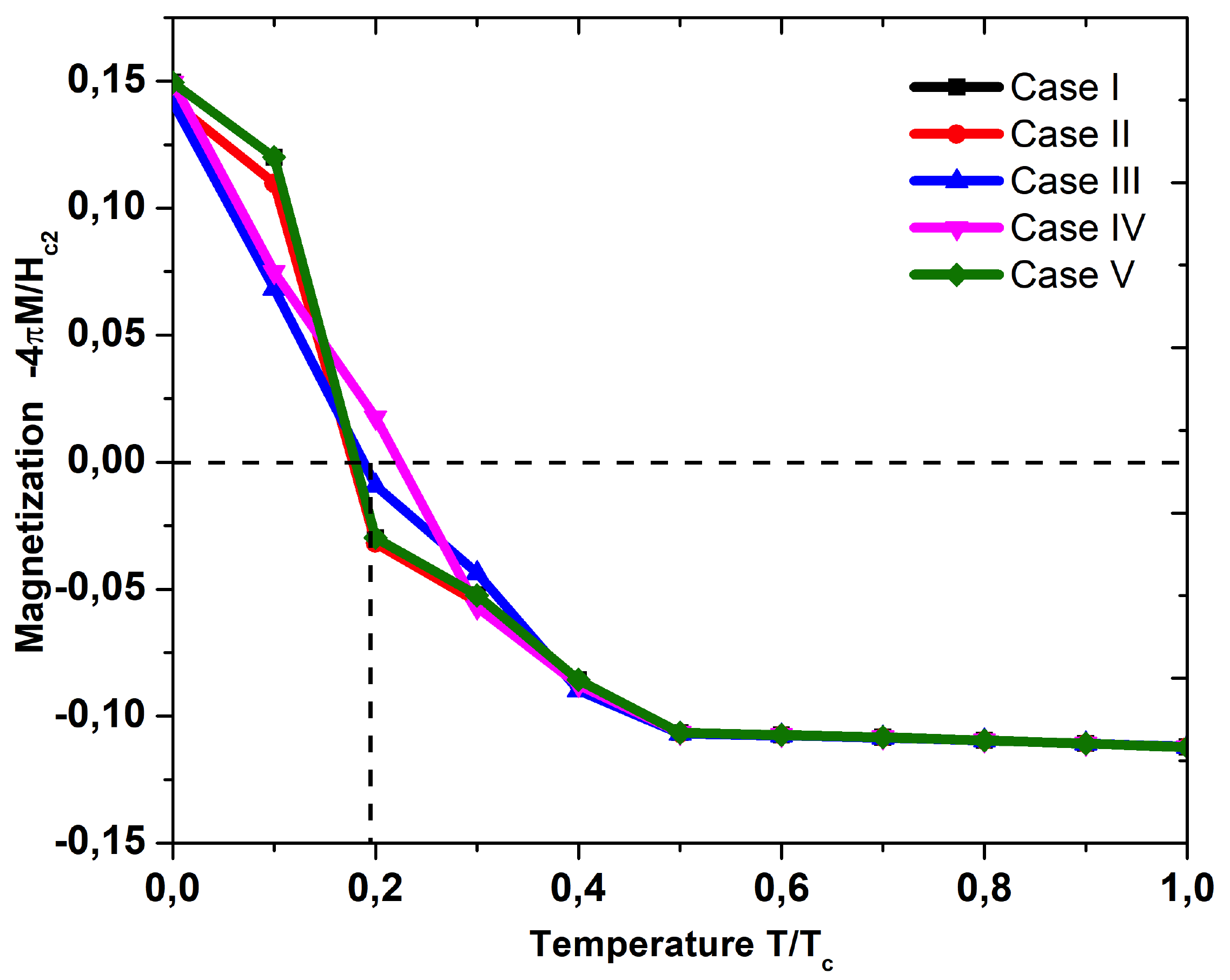}
\caption{Magnetization $-4\pi\mathbf{M}/H_{c2}$ as function of the temperature $T$, at $H=1.1$ in ZFC process for the cases I, II, III, IV and V.}
\label{Magnetization}
\end{figure}

\begin{figure}[htbp!]
\centering
\includegraphics[scale=0.3]{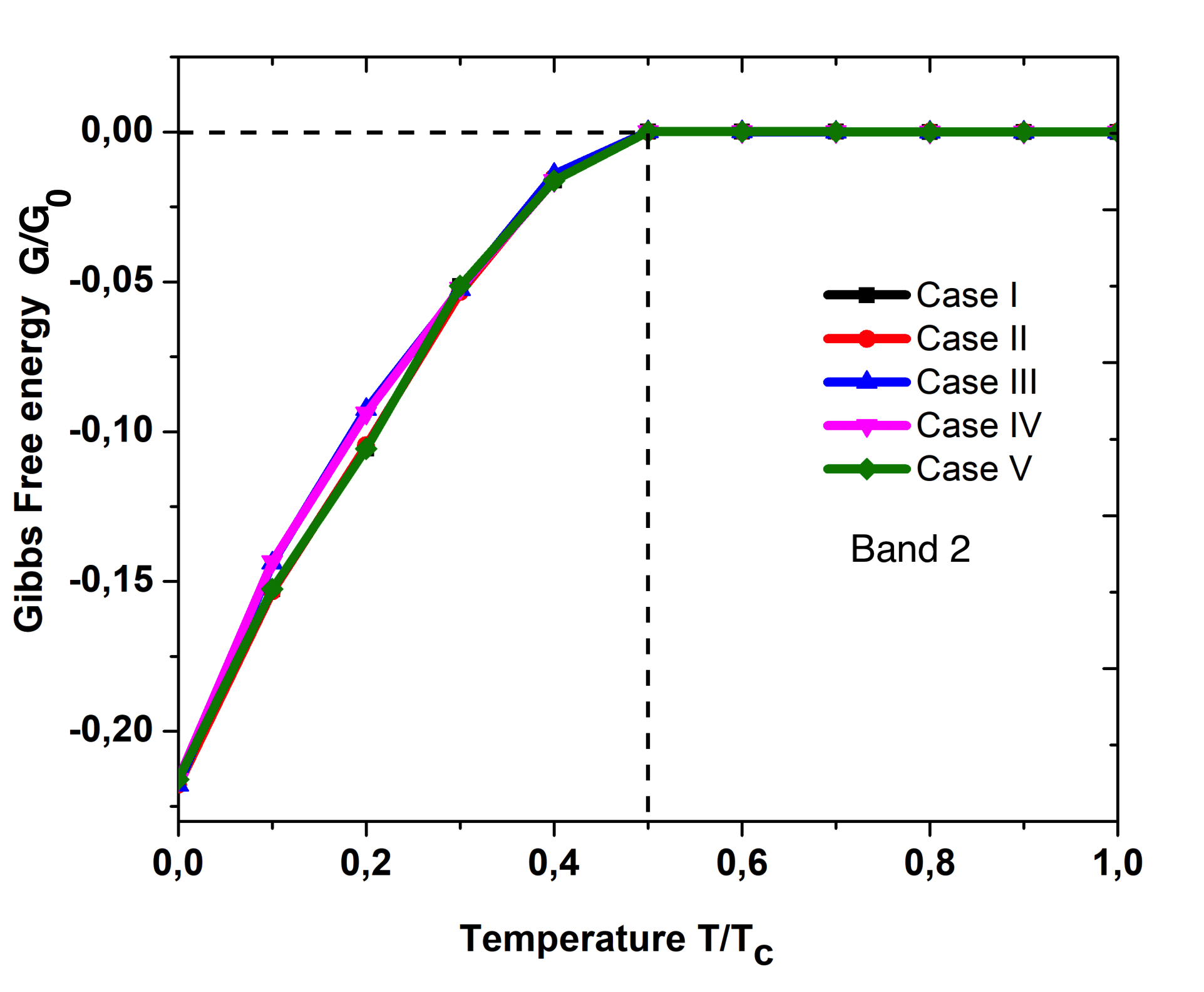}
\caption{Gibbs free energy in function of the temperature $G(T)$, for ZFC at $H=1.1$, for  the band 2 for the cases I, II, III, IV, and V.} 
\label{Gibbspsi2}
\end{figure}

\begin{figure}[htbp!]
\centering
    \includegraphics[scale=0.22]{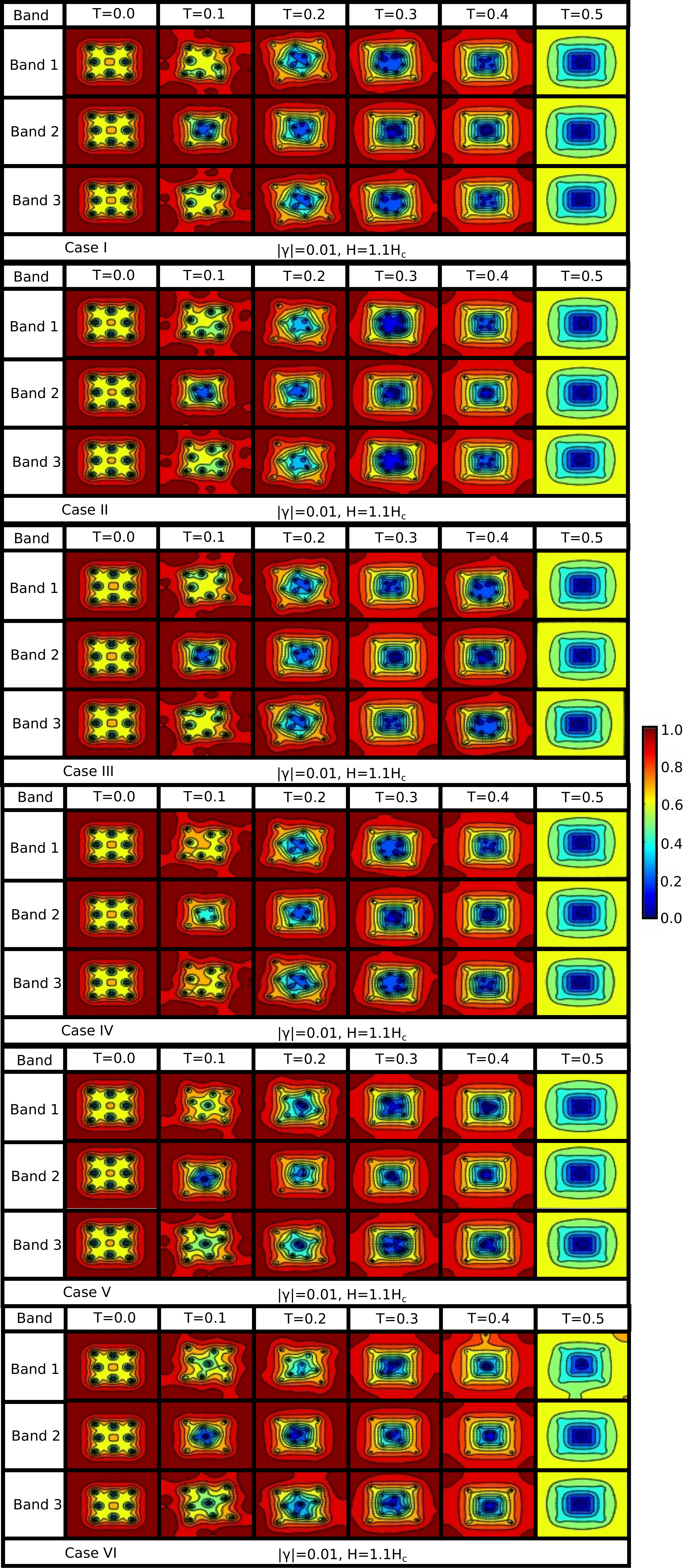}
    \caption{Vortex configuration in the ZFC process at $H=1.1$ for different temperatures $T$ and for all cases presented in Tab. \ref{Table1}, for the bands 1, 2 and 3.}
    \label{case1a}
\end{figure}

\begin{figure}[htbp!]
   \centering
    \includegraphics[scale=0.23]{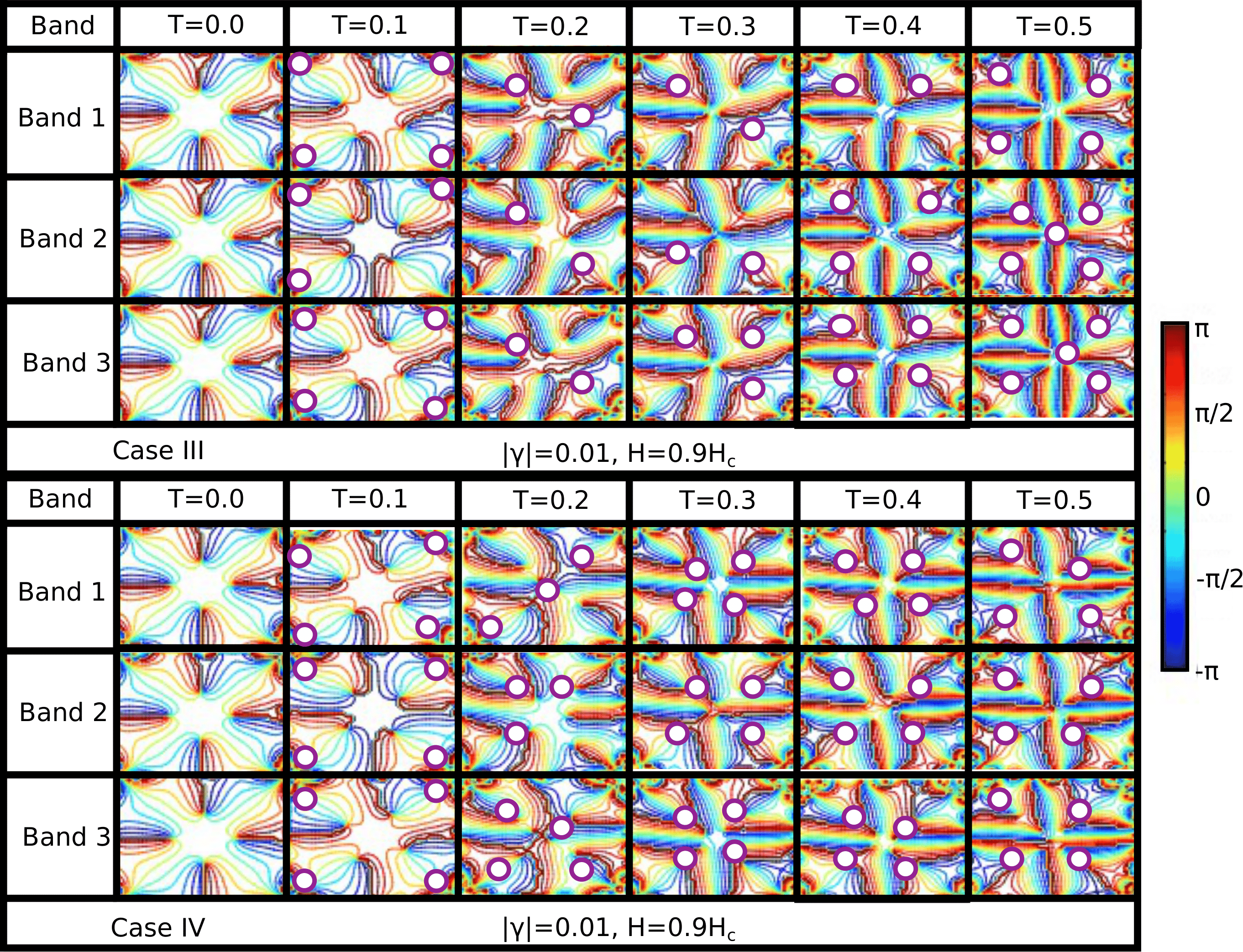}
   \caption{Phase $\theta$ of the order parameter for the band 1, 2 and 3 for the cases III and IV, in the ZFC process at several $T$  and $H=0.9$. The anti-vortex position is indicated by the symbol $\circ$. Dark and bright regions represent values of the modulus of the order parameter (as well as, $\theta/2\pi$) from 0 to 1.}
    \label{phase1a}
\end{figure}

\begin{figure}[htbp!]
\centering
\includegraphics[scale=0.3]{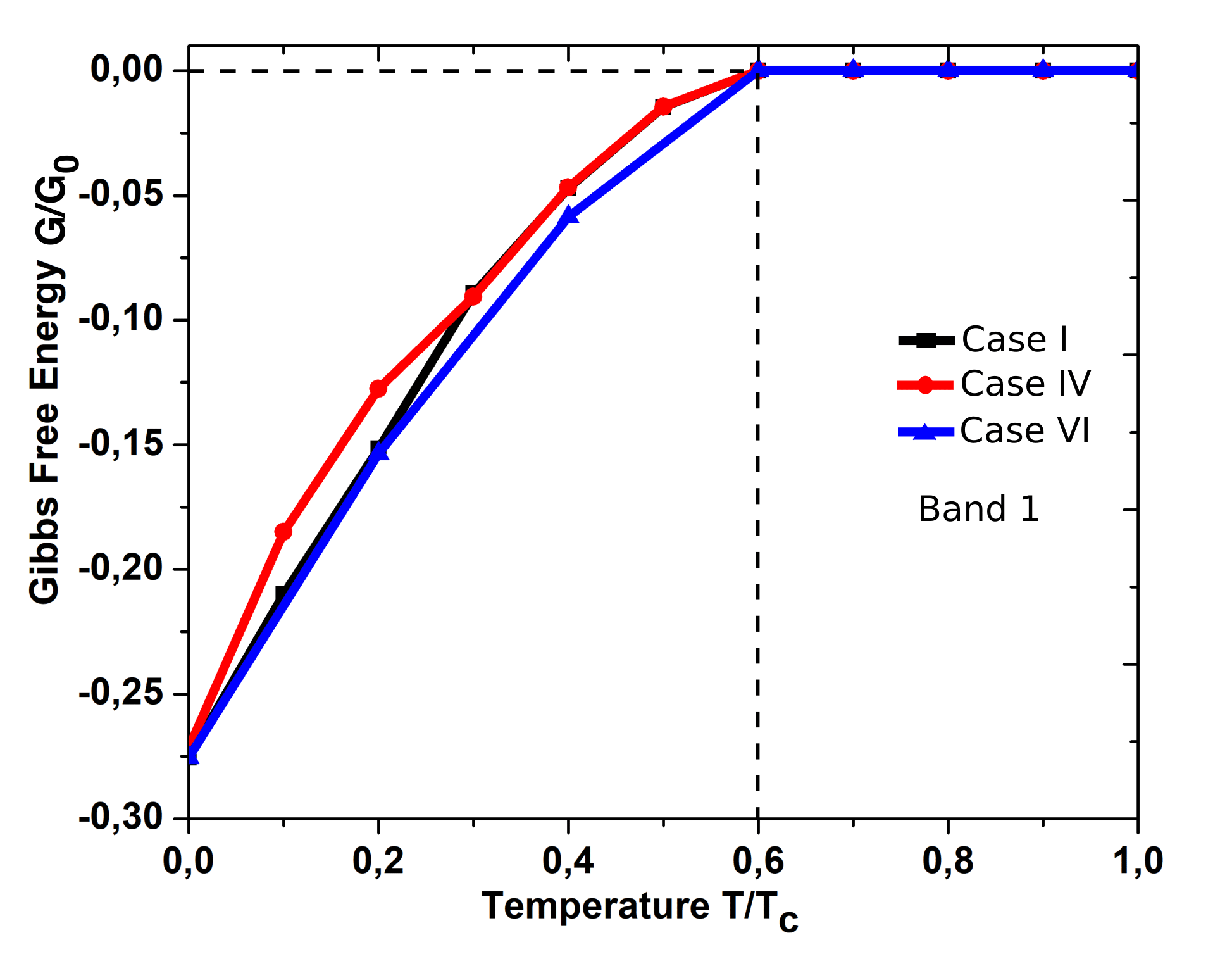}
\caption{Gibbs free energy  $G$ in function of $T$ for the order parameter $\psi_{1}$ in a  ZFC process at $H=0.9$.} 
\label{GibbH1}
\end{figure}

\begin{figure}[htbp!]
\centering
\includegraphics[scale=0.3]{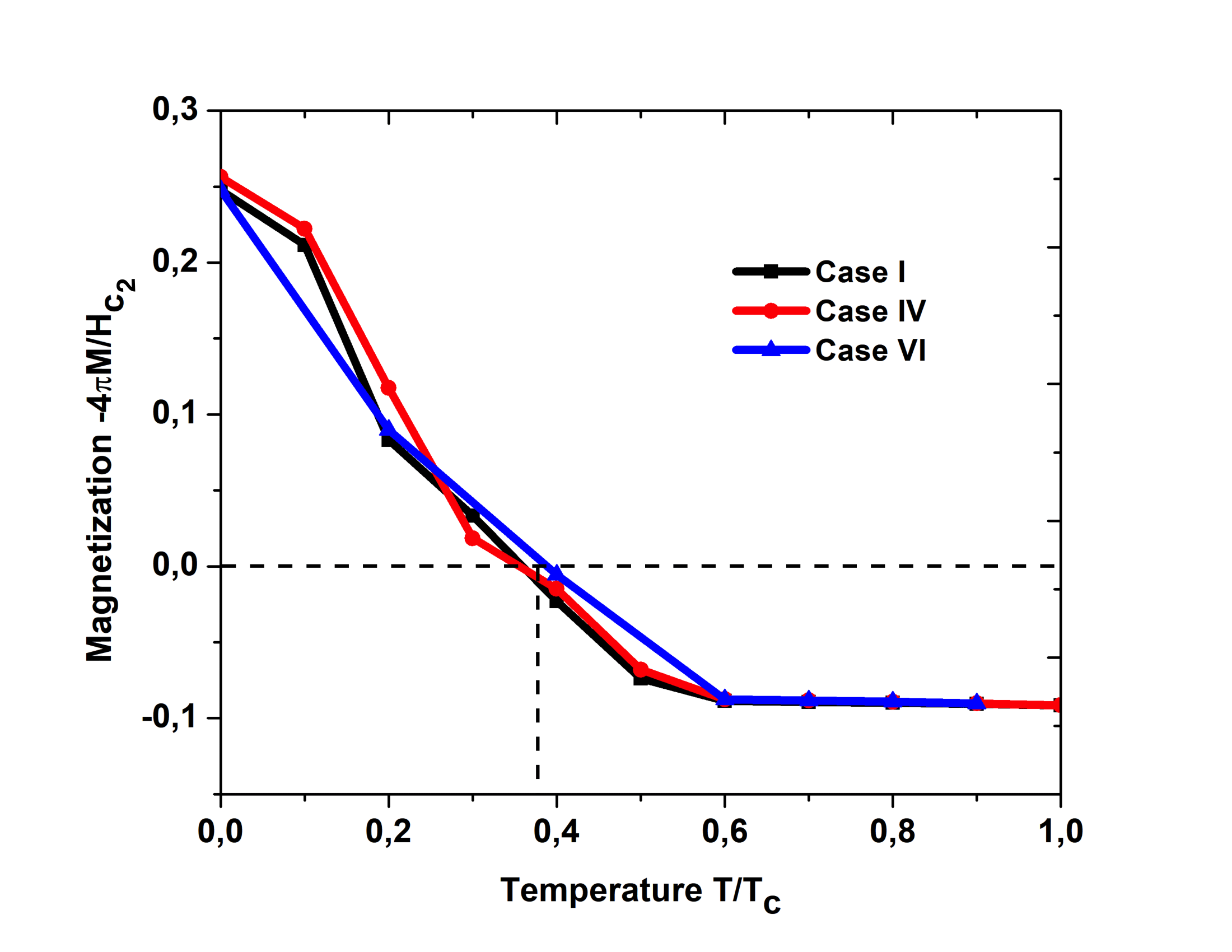}
\caption{Magnetization in function of the temperature $T$ for the first order parameter $\psi_{1}$ in ZFC with $H<H_{1}$.} 
\label{MagH1}
\end{figure}

With this, we begin to present the ZFC process, for two chosen values of the external field $H=1.1$ and $H=0.9$. We show the magnetization $-4\pi M$as a function of temperature $T$, for the superconducting sample in Fig. \ref{Magnetization} for the cases I, II, II, IV and V show in the Table \ref{Table1}. There are some properties in this graph, on which we call our attention, the functional form in general is in accordance with what is expected in a conventional one-band system; where we observe changes in the slope at various points, which is indicative of the entry of vortices into the sample but there are positives values of this magnetization, indicating an unconventional para-magnetic behavior. This behaviour is due to the fact that in general the competition effect in the superconducting bands generates frustration in the general superconducting state, due to the different forms of oscillation in the crystal lattice. i.e, because each of the bands presents a superconducting gap, the network in its oscillations can generate increases in that energies in each of the bands while decreasing in others, this is generally due to the inter-band coupling that has been taken and a non-monotonic interaction between the vortices present in each of the bands, which due to the super-current, generates a long-range attraction, but with a short-range repulsion, this effect is one of those responsible for the creation of the clusters of vortices in multi-band mesoscopic samples. 
Additionally, we observe the effect that the variations of the parameters $\hat{\gamma}$ and $\zeta$ has on the magnetization, because this magnetization is a macroscopic variable, we only observe that the value of the magnetic field in which one or more vortices penetrates to sample ($H_{1}$), is different for each of the cases, at different temperatures.
\begin{figure}[htbp!]
\centering
\includegraphics[scale=0.3]{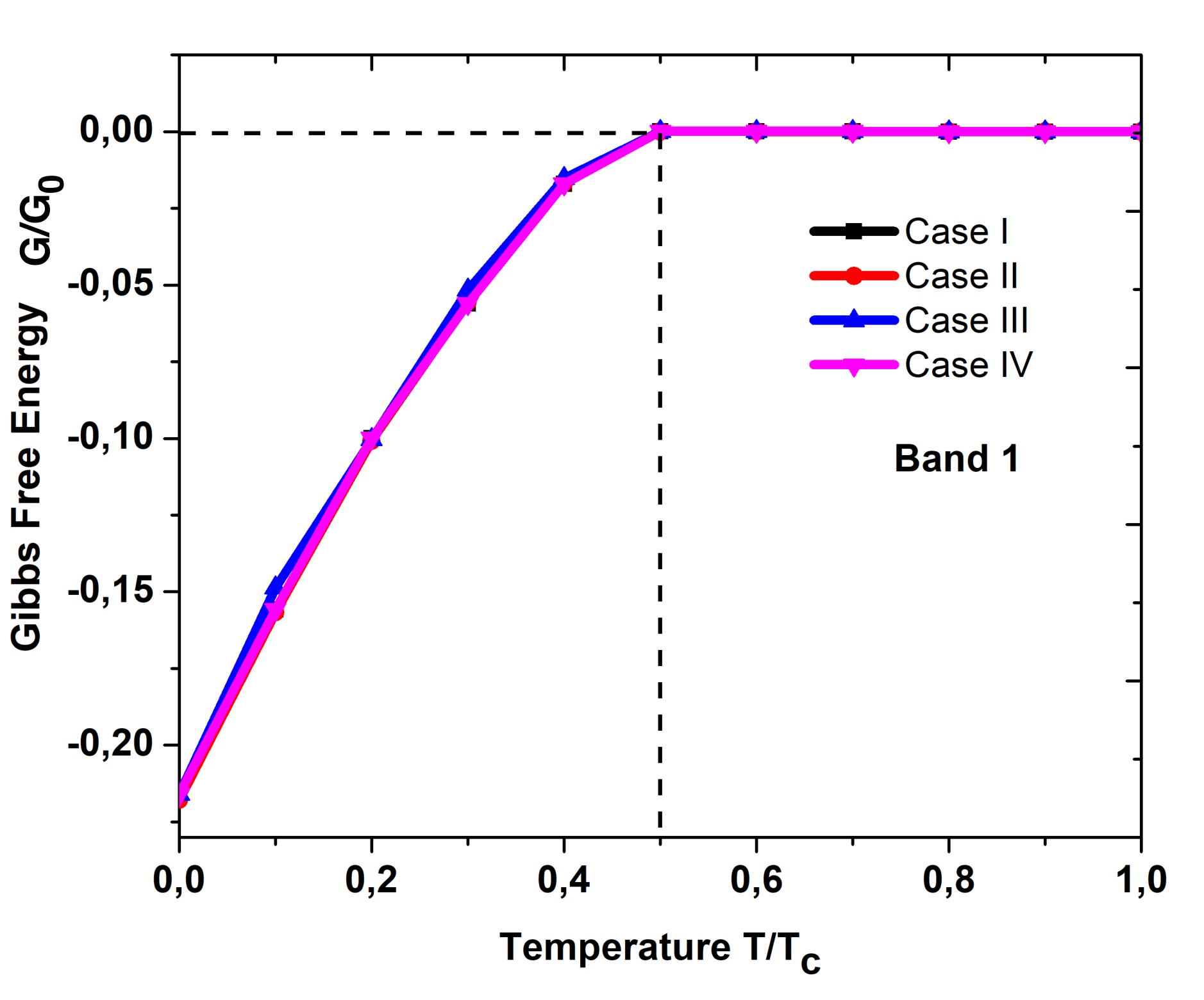}
\caption{Gibbs free energy $G(T)$, for the band 1 and the cases I, II, III and IV.} 
\label{Gibbspsi1}
\end{figure}
Now in Fig. \ref{Gibbspsi1}, we plot the Gibbs energy density as a function of the temperature for the band 1 or first condensate $\psi_{1}$. We observe that as the temperature is varied, the system obtains a minimization of the energy and for values $T>0.5T_{c}$, the sample is in a normal state, with this, additional in Fig. \ref{Gibbspsi2}, we observe the Gibbs free energy, for the band 2 or second order parameter $\psi_ {2}$ (the third condensate has the same behavior). Now, we observe that in both cases Figs. \ref{Gibbspsi1}-\ref{Gibbspsi2}, the behavior is very close to the conventional one for the ZFC process, that is, the functional density tends to zero, as the temperature increases and the system transitions to a normal state; although it is observed that in Fig. \ref{Gibbspsi2}, there is first the entry of vortices and even more, the amount of vortices for each of the different temperatures is different, as we will present below.
Thus, we proceed in Fig. \ref{case1a}, we present the super-electron density in a logarithmic scale  $\ln|\psi_{i}| ^{2}$, with $ i = 1,2,3 $ at different $T$, at $H=1.1$, in the ZFC process, for the six cases presented in Tab. \ref{Table1}. It is important to note that in Fig. \ref{case1a} for $T=0.0$, the vortices are in the same position and the global energy in the identical system, see Fig. \ref{Gibbspsi1} and Fig. \ref{Gibbspsi2}. Now, as the temperature increase, the vortex configuration and the number of them, in each band change; the repulsive force between the vortices decrease, due to the tunneling effect between the bands, and they conglomerate in clusters, near the center of the sample (centre-symmetric superconductor). Additionally, as the temperature increases, it breaks said coherence in the phases and by establishing that each of the phonons for each of the superconducting gaps present different energy, the oscillations in phase account for non-monotonic interactions in the system, leading to the tunneling of Cooper pairs between bands, exists and with said tunneling the interaction between the bands, presents an attractive interaction in the long range and repulsive in the short, giving the appearance of a cluster configuration in the vortices, which rotate according to a central position of high symmetry.
Now in the Fig. \ref{phase1a}, we present the phase of the order parameter, for each of the condensates at several $T$, we observe the creation of anti-vortices in the bands (the anti-vortex position is indicated by the symbol $\circ$), because the sample is homogeneous, the only indication of the phenomenon that generates this creation in one of the bands, is due to the tunneling of an anti-vortex in another of the superconducting bands, with which we have marked in the fig, the band from which said tunneling was created and in which it entered. Additionally, the creation of several anti-vortices is evident, with which the tunneling for said coupling generates different vortex anti-vortex states, with different activation energies, which is reflected in the interaction in all bands.
In the Fig. \ref{GibbH1} and the Fig. \ref{MagH1}, we present the Gibbs free energy density and magnetization, for what we call case II, i.e.,  the ZFC process with a fixed external magnetic field of $H<0.9$, we observe that for values $T>0.5$, the system transitions to a normal state, in both figures, only $3$ particular cases have been presented, which minimize the functional energy, due to the chirality observed in the system, not for all cases such functional is minimized. Additionally, we observe that in all cases the value of $H_{1}$ is different, this establishes that the value of $\gamma$ generates that the system allows entering for lower values of the temperature, more quickly the vortices and generates transient states for lower energy values.

\begin{figure}[htbp!]
   \centering
    \includegraphics[scale=0.25]{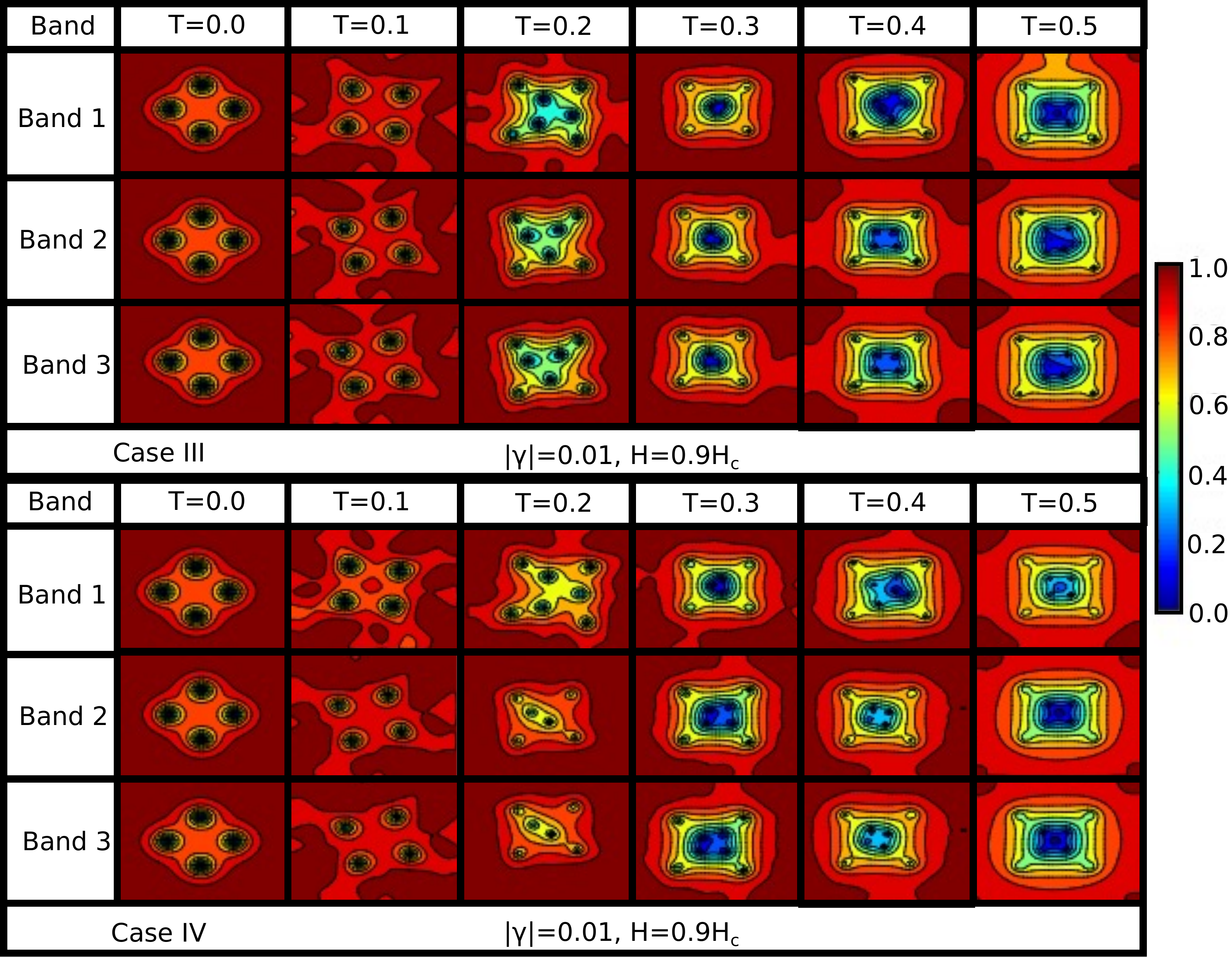}
    \caption{Vortex configuration in the ZFC process at $H=0.9$ for different  $T$, and for the III and IV cases presented in Tab. \ref{Table1}. }
    \label{case3h09ZFC}
\end{figure}

\begin{figure}
   \centering
    \includegraphics[scale=0.22]{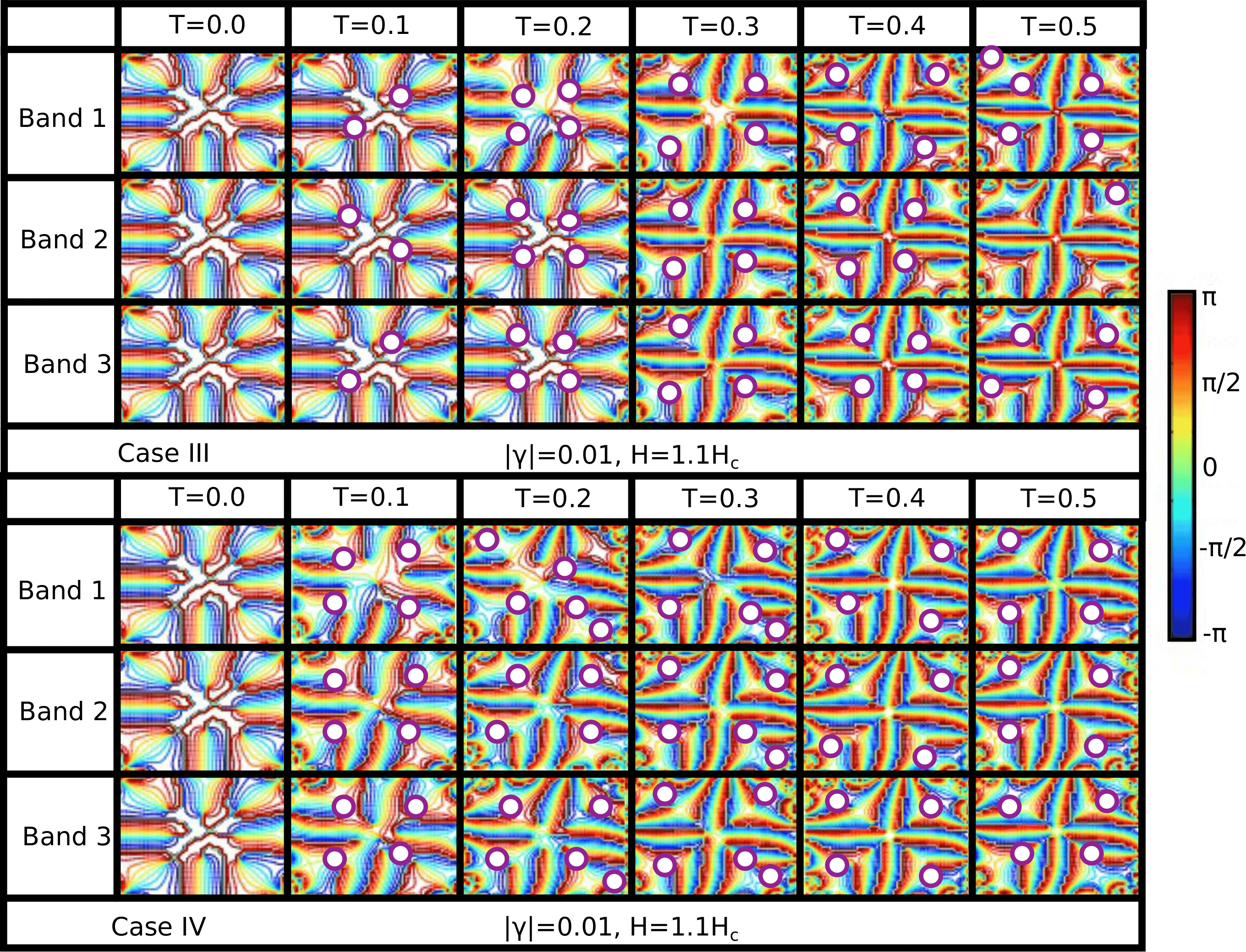}
    \caption{Phase $\theta$ of the order parameter at several $T$ for the cases III and IV for $H=1.1$ in the ZFC process. The anti-vortex position is indicated by the symbol $\circ$. Dark and bright regions represent values of the modulus of the order parameter (as well as, $\theta/2\pi$) from 0 to 1.}
    \label{phaseh09ZFC}
\end{figure}

Now in Fig. \ref{case3h09ZFC}, we present the vortex state in the ZFC process at $H=0.9$, where a quantity and configuration of vortices are initially observed for each of the condensates. Important, to highlight in Fig. \ref{case3h09ZFC}, that for the different $\hat{\gamma}$ and $\zeta$, there are several vortex configurations in meta-stables states in the sample. For this vortex configurations, the system cannot to minimize the Gibbs free energy, that is; energetically speaking, the sample tends to generate a normal-state more quickly than in the other configurations and this generates that a transient-state are more varied before obtaining the stable-state.
Finally, before approaching the system in process FC, we observe in Fig. \ref{phaseh09ZFC}, the phase of the order parameter, for the case III and case IV, we have labeled with a white symbol where the position of anti-vortex in the bands is observed. 
\subsection{Field-Cooling process at $T$ constant.}\label{Section3}
\begin{figure}[H]
   \centering
    \includegraphics[scale=0.25]{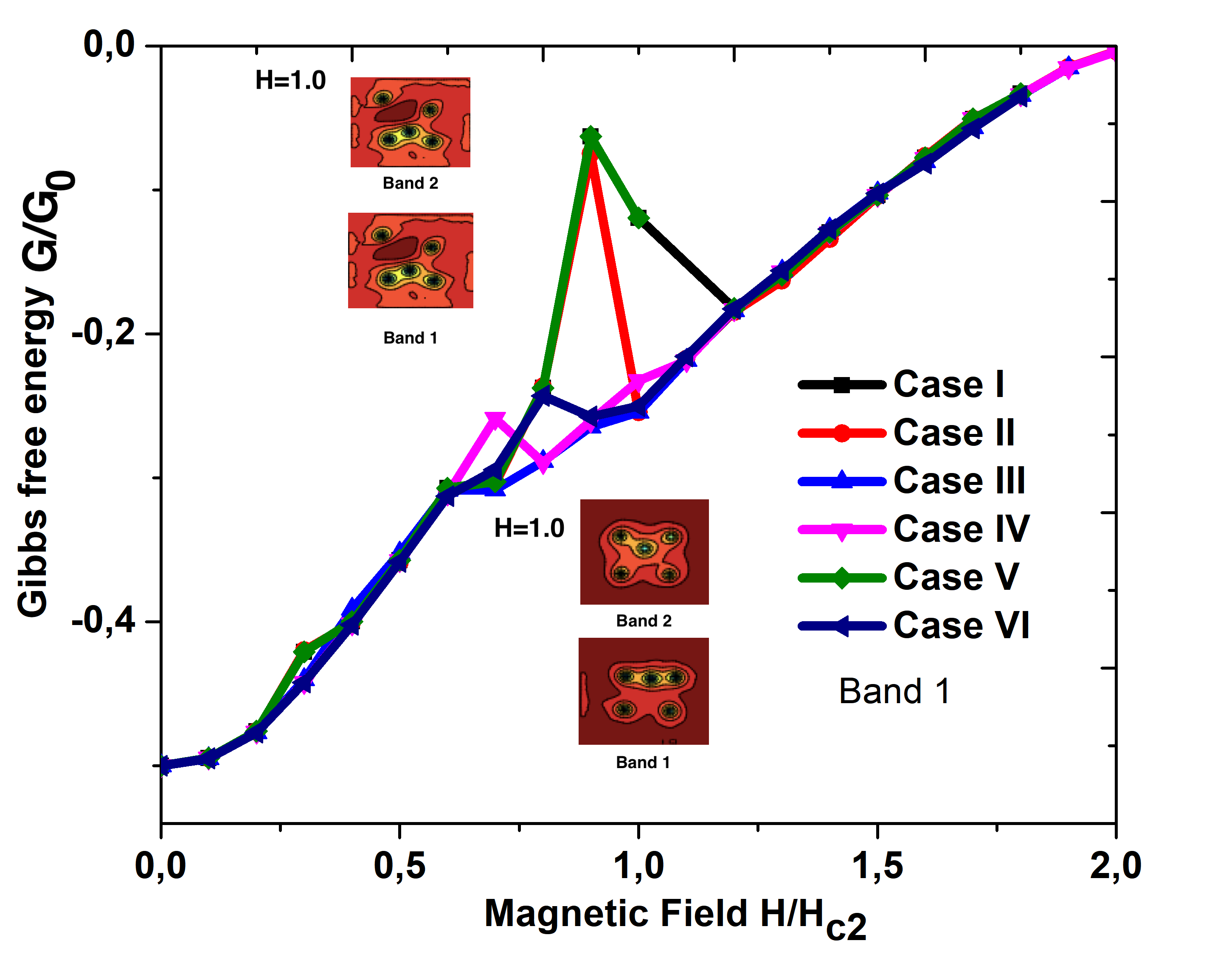}
    \caption{Gibbs free energy, $G(H)$,  for the studied cases (show in the Tab. \ref{Table1}). (Inset) Vortex state for the band 1 and the band 2, at $H=1.0$ for the case II (up) and for the case V (down).}
    \label{EnergyFC}
\end{figure}

\begin{figure}[H]
   \centering
    \includegraphics[scale=0.25]{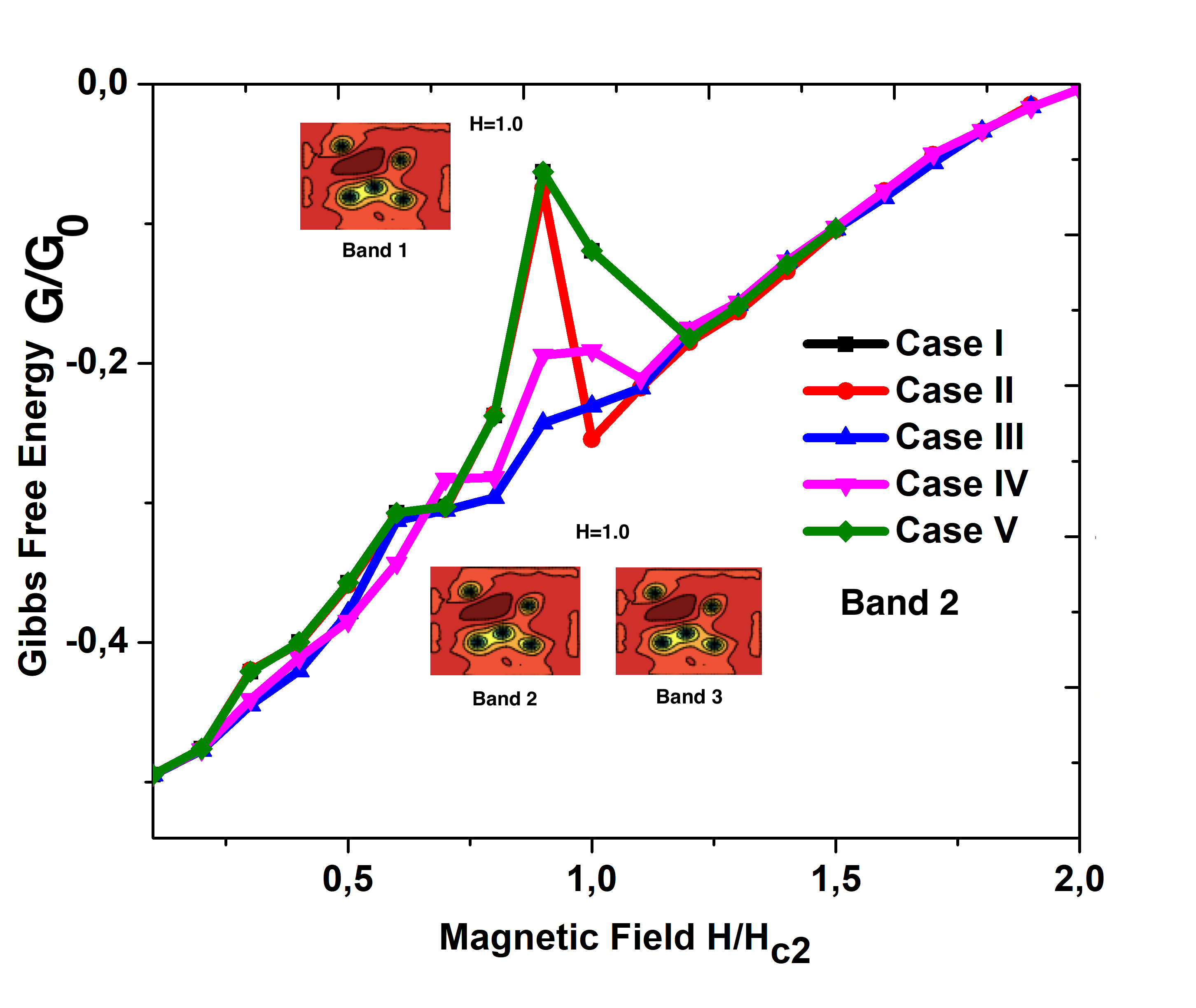}
    \caption{Gibbs free energy $G(H)$, for the studied cases (show in the Tab. (\ref{Table1})), at $H=1.0$ for the all bands for the case  II.}
    \label{EnergyFC2}
\end{figure}

\begin{figure}[H]
   \centering
    \includegraphics[scale=0.28]{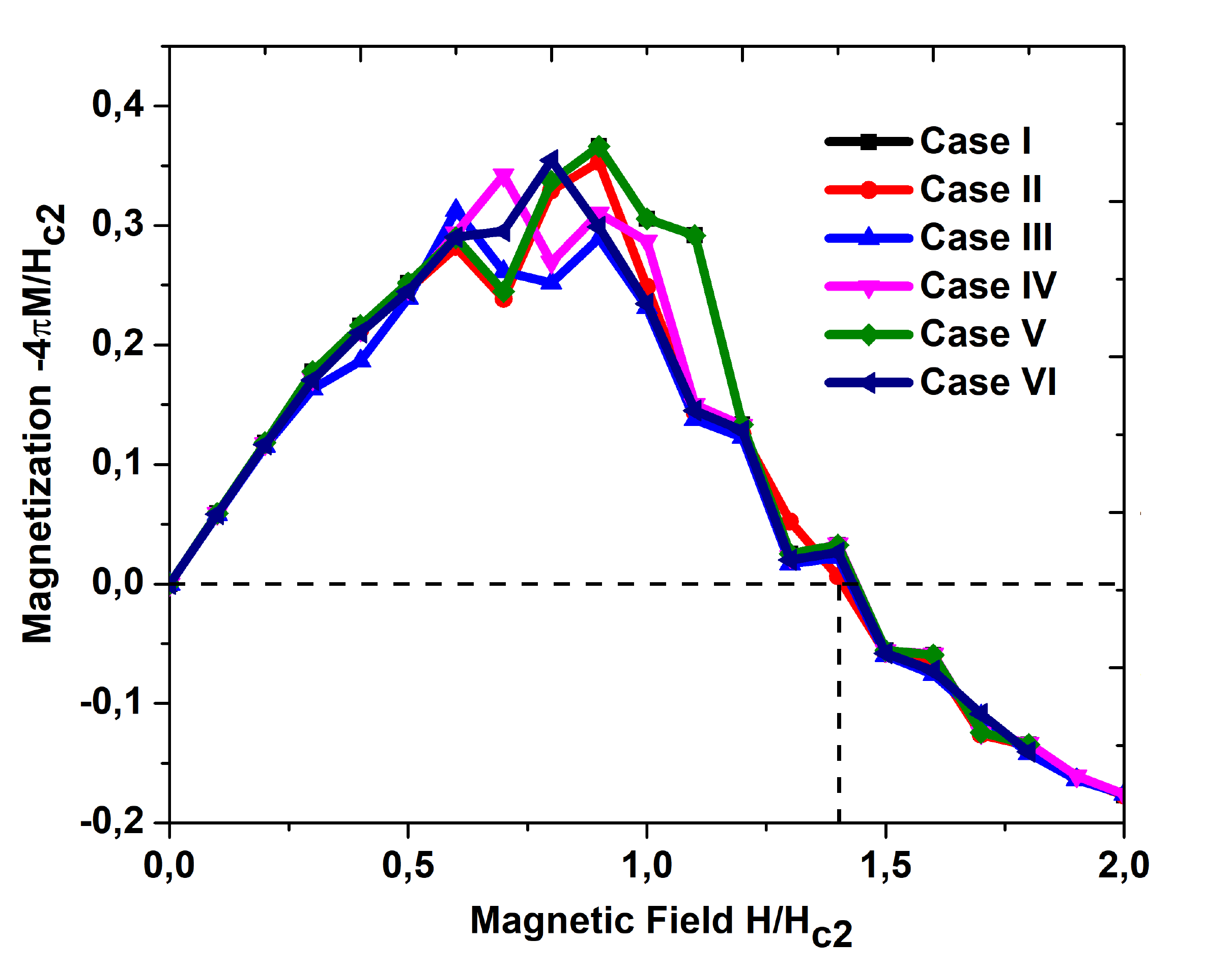}
    \caption{Magnetization as function of the external magnetic field $H$, for the studied cases (showed in Tab.(\ref{Table1}).}
    \label{MagnetFC}
\end{figure}

Now, our discussion will focus on the study of the three-band system submitted to FC process, we choose $T=0$ for all simulations for this case. In the n Fig. \ref{EnergyFC}, we present the Gibbs free energy density $G$ as a magnetic field function at $T=0$, for the the cases I, II, III, IV and V (shown Tab. \ref{Table1}). We observe a soft behaviour of $G(H)$ for $H\leq 0.65$ and $H\geq 1.22$, however, we call attention to the existence of a particular interval $0.65\leq H\leq 1.22$ in which a very appreciable variation in energy is observed. We present the  vortex state for the band 1 and the band 2 at $H=1.0$ , we observe the creation of an anti-vortex created in one band and tunneling to the two alternate band see(purple dots in Fig.\ref{phaseh09ZFC}). 
In the Fig. \ref{EnergyFC2} we present the Gibbs free energy curve for the band 2, and (inset) the vortex state for all bands at $H=1.0$ for the case II.  We observe that the vortex configuration is different for the cases studied in the insets of the Fig. \ref{EnergyFC}.
With this, in the Fig. \ref{MagnetFC}, we present the magnetization curve as a magnetic field function for the band 1 and all studied cases. It is extremely important to establish that for $H>1.45$, the sample presents a para-magnetic behavior $\mathbf{M}>0$, we think that this effect is due to the tunneling of vortex and anti-vortex between the bands, as it is generated in one of the bands and transported to another, it generates interactions between the Cooper pairs of the band in which enter and the tunneling that give out-of-phase oscillations between said Cooper pairs, for this reason we have this anomalous behavior in the magnetization curve..
In the Fig. \ref{FC1} we present the vortex state for the bands 1, 2 and 3, at $H=0.6;0.7;0.8;0.9;1.0;1.1;1.2;1.3;1.4$, for the case I, II, III and IV at $T=0$. We observe that as $H$ increases, the vortex configuration is different.
\begin{widetext}
\begin{center}
\begin{figure}[H]
   \centering
    \includegraphics[scale=0.33]{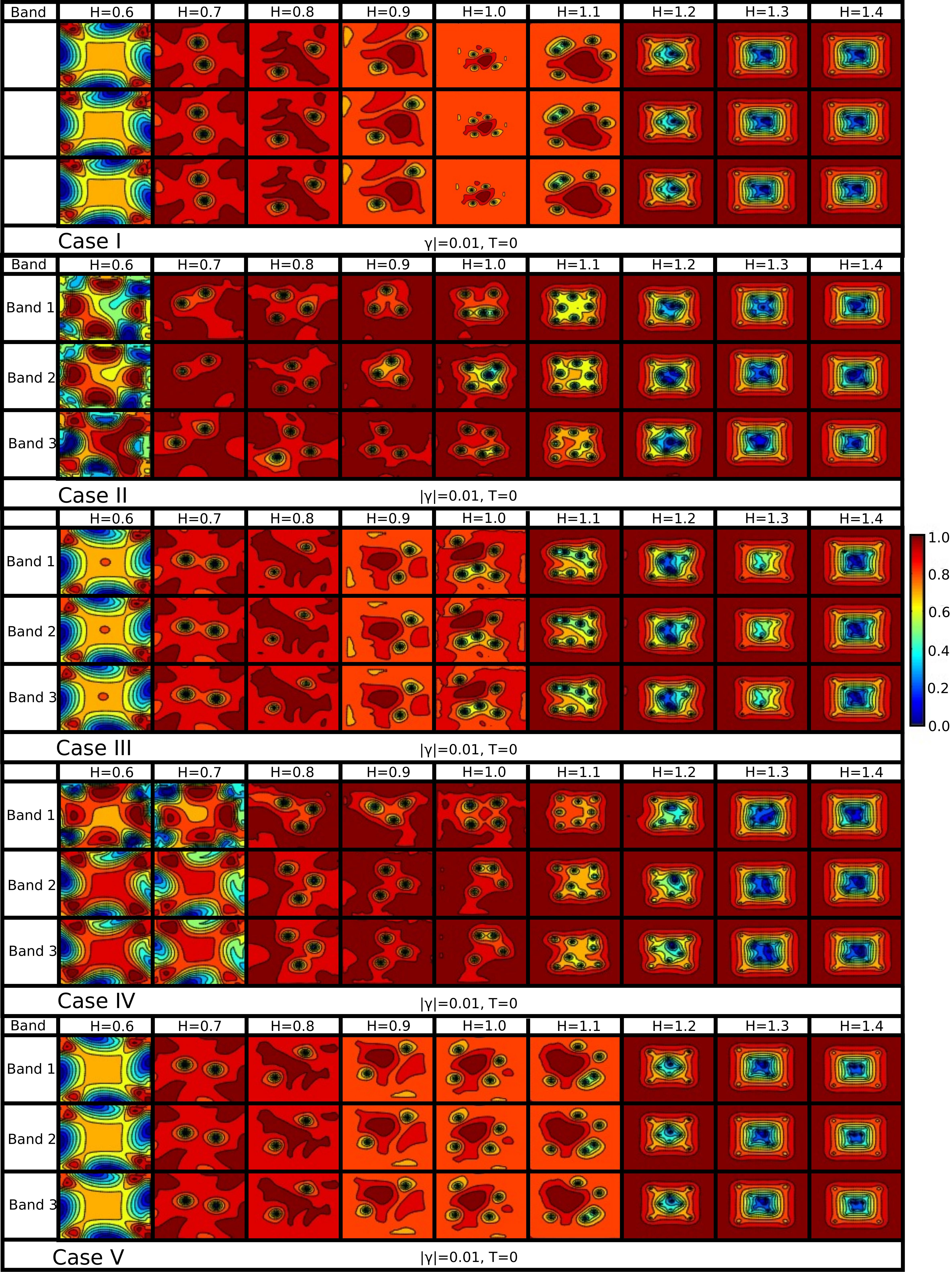}
    \caption{Vortex configuration for the bands 1, 2 and 3, for the cases the case I, II, III and IV (presented in Tab. \ref{Table1}), at indicates $H$.}
    \label{FC1}
\end{figure}
\end{center}
\end{widetext}
\section{Conclusions}\label{Section3}
In the present work, we have studied superconducting properties in a three-band nano-prism, composed of three order parameters that interact with each other through an inter-parameter Josephson-type coupling. In this three-dimensional-orbital system we have studied the magnetization, superconducting electronic density and the Gibbs free energy, under  Zero-Field-Cooling and Field-Cooling processes. In this three-band system, an interesting paramagnetic behaviour have been found that are not observed in a system with a single or two (conventional) orbitals, we attributed this behavior to the competition between bands, effect similar to that obtained in magnetic frustration. We have also observed the tunneling of vortex and anti-vortex between the superconducting bands as well as vortex clusters. We propose that the study of these properties can be used to analyze three-orbital superconducting systems such as copper oxides \ce{CoO2} (Cuprates), iron oxides \ce{FeO2} (Iron pnictides) and nickel oxides \ce{NiO2} (Niquelates).
\section*{ACKNOWLEDGMENTS}
C. A. Aguirre, would like to thank the Brazilian agency CAPES, for financial support and the Ph.D, Grant number: 0.89.229.701-89. J. Faúndez and S. G. Magalhães thank FAPERGS, CAPES and CNPq for partially financing this work under the Grant PRONEX 16/0490-0.
$^{\dagger}$email: \href{cristian@fisica.ufmt.br}{cristian@fisica.ufmt.br}

\newpage
\begin{widetext}
\appendix
\section{Josephson coupling }\label{apendice1}

From Gibbs energy density, Eq. (\ref{Gibbs1}), we will consider the Josephson coupling for three order parameters thought the isotropic inter-band coupling $\gamma$ as,

\begin{equation}
\gamma (\psi_1^*\psi_2\psi_3+\psi_1 \psi_2^* \psi_3^*)+ \gamma (\psi_1\psi_2^*\psi_3+\psi_1^{*} \psi_2 \psi_3^*)+ \gamma (\psi_1\psi_2\psi_3^*+\psi_1^* \psi_2^*\psi_3)
\label{A1}
\end{equation}

where the real order parameter is defined as $\psi_{i}=|\psi_{i}|e^{i\theta_{i}}$ and it conjugate $\psi^{*}_{i}=|\psi_{i}|e^{-i\theta_{i}}$. Them, replacing in the firth part of the Eq. (\ref{A1}) we have

\begin{eqnarray*}
\gamma (\psi_1^*\psi_2\psi_3+\psi_1 \psi_2^*\psi_3^*)=\gamma (|\psi_{1}||\psi_{2}||\psi_{3}|e^{i(\theta_{3}+\theta_{2}-\theta_{1})}+ |\psi_{1}||\psi_{2}||\psi_{3}|e^{-i(\theta_{3}+\theta_{2}-\theta_{1})})
\end{eqnarray*}
\begin{eqnarray*}
=2\gamma |\psi_{1}||\psi_{2}||\psi_{3}|\cos(\theta_{3}+\theta_{2}-\theta_{1}).
\end{eqnarray*}

Now, in similar way, we can write the second part of Eq. (\ref{A1}) as
\begin{eqnarray*}
\gamma (\psi_1\psi_2^* \psi_3+\psi_1^* \psi_2 \psi_3)=\gamma (|\psi_{1}||\psi_{2}||\psi_{3}|e^{i(\theta_{3}-\theta_{2}+\theta_{1})}+ |\psi_{1}||\psi_{2}||\psi_{3}|e^{-i(\theta_{3}-\theta_{2}+\theta_{1})}),
\end{eqnarray*}
\begin{eqnarray*}
=2\gamma |\psi_{1}||\psi_{2}||\psi_{3}|\cos(\theta_{3}-\theta_{2}+\theta_{1})
\end{eqnarray*}
and for the third part 
\begin{eqnarray*}
\gamma (\psi_1\psi_2 \psi_3^*+\psi_1^* \psi_2^* \psi_3)=\gamma (|\psi_{1}||\psi_{2}||\psi_{3}|e^{i(-\theta_{3}+\theta_{2}+\theta_{1})}+ |\psi_{1}||\psi_{2}||\psi_{3}|e^{-i(-\theta_{3}+\theta_{2}+\theta_{1})}),
\end{eqnarray*}
\begin{eqnarray*}
\gamma (\psi_1\psi_2 \psi_3^*+\psi_1^* \psi_2^* \psi_3)=2\gamma |\psi_{1}||\psi_{2}||\psi_{3}|\cos(-\theta_{3}+\theta_{2}+\theta_{1})
\end{eqnarray*}

thus, the Eq (\ref{A1}) can be write as

\begin{equation}
    2\gamma |\psi_{1}||\psi_{2}||\psi_{3}|\cos(\theta_{3}+\theta_{2}-\theta_{1})+2\gamma |\psi_{1}||\psi_{2}||\psi_{3}|\cos(\theta_{3}-\theta_{2}+\theta_{1})+2\gamma |\psi_{1}||\psi_{2}||\psi_{3}|\cos(\theta_{1}+\theta_{2}-\theta_{3})
    \label{A2}.
\end{equation}

\begin{itemize}
    \item The first functional derivative, $\delta\psi^{*}_{1}=\frac{\partial}{\partial\psi^{*}_{1}}$

{\small\begin{eqnarray}
2\gamma |\psi_{2}||\psi_{3}|\Big[\delta {\psi_{1}^{*}} [|\psi_{1}|\cos(\theta_{3}+\theta_{2}-\theta_{1})]+\delta {\psi_{1}^{*}} [|\psi_{1}|\cos(\theta_{3}-\theta_{2}+\theta_{1})]+\delta {\psi_{1}^{*}} [|\psi_{1}|\cos(\theta_{1}+\theta_{2}-\theta_{3})]   \Big]
\label{f1}
\end{eqnarray}}

for the firth part
\begin{equation*}
\frac{\partial}{\partial \psi_{1}^{*}} \Big[|\psi_{1}|\cos(\theta_{3}+\theta_{2}-\theta_{1})\Big] =   |\psi_{1}|\Big[-\sin( \theta_{3}+\theta_{2}-\theta_{1}) \cdot -(\frac{\partial\theta_{1}}{\partial\psi_{1}^{\dagger}}) \Big]+\cos(\theta_{3}+\theta_{2}-\theta_{1})\Big[  \frac{\psi_{1}}{2|\psi_{1}|} \Big],
\end{equation*}
\begin{equation*}
\frac{\partial}{\partial \psi_{1}^{*}} \Big[|\psi_{1}|\cos(\theta_{3}+\theta_{2}-\theta_{1})\Big] =   |\psi_{1}|\sin( \theta_{3}+\theta_{2}-\theta_{1})  (\frac{\partial\theta_{1}}{\partial\psi_{1}^{\dagger}}) + \frac{\psi_{1}}{2|\psi_{1}|} \cos(\theta_{3}+\theta_{2}-\theta_{1})
\end{equation*}
and with the phases, see the appendix (\ref{apendiceB}), we can write the last Eq. as
\begin{equation}
\frac{\partial}{\partial \psi_{1}^{*}} \Big[|\psi_{1}|\cos(\theta_{3}+\theta_{2}-\theta_{1})\Big] =  i|\psi_{1}|\frac{\psi_{1}}{2|\psi_{1}|^{2}}  \sin( \theta_{3}+\theta_{2}-\theta_{1})  + \frac{\psi_{1}}{2|\psi_{1}|} \cos(\theta_{3}+\theta_{2}-\theta_{1})
\label{A3}
\end{equation}

for the second term

\begin{equation*}
    \delta\psi_{1}^{*}[|\psi_{1}|\cos(\theta_{3}-\theta_{2}+\theta_{1})]=|\psi_{1}|\cdot -\sin(\theta_{3}-\theta_{2}+\theta_{1})\left( \frac{\partial\theta_{1} }{\partial\psi_{1}^{*}}\right)+ \cos( \theta_{3}-\theta_{2}+\theta_{1}) \frac{\psi_{1}}{2|\psi_{1}|},
\end{equation*}
\begin{equation}
    \delta\psi_{1}^{*}[|\psi_{1}|\cos(\theta_{3}-\theta_{2}+\theta_{1})]=-\frac{\psi_{1}}{2|\psi|^{2}}\cdot \sin(\theta_{3}-\theta_{2}+\theta_{1})+ \frac{\psi_{1}}{2|\psi_{1}|}\cos( \theta_{3}-\theta_{2}+\theta_{1}) 
    \label{A5}
\end{equation}

and for the third term
\begin{equation*}
    \delta\psi_{1}^{*}[|\psi_{1}|\cos(\theta_{1}+\theta_{2}-\theta_{3})]=|\psi_{1}|\cdot -\sin(\theta_{1}+\theta_{2}-\theta_{3})\left( \frac{\partial\theta_{1} }{\partial\psi_{1}^{*}}\right)+ \cos( \theta_{1}+\theta_{2}-\theta_{1}) \frac{\psi_{1}}{2|\psi_{1}|},
\end{equation*}
\begin{equation}
    \delta\psi_{1}^{*}[|\psi_{1}|\cos(\theta_{1}+\theta_{2}-\theta_{3})]=-\frac{\psi_{1}}{2|\psi|^{2}}\cdot \sin(\theta_{1}+\theta_{2}-\theta_{1})+ \frac{\psi_{1}}{2|\psi_{1}|}\cos( \theta_{1}+\theta_{2}-\theta_{3}) 
    \label{A7}
\end{equation}

Now, grouping the Eqs (\ref{A3}), (\ref{A5}) and (\ref{A7}), we have that

\begin{multline}
    i|\psi_{1}|\frac{\psi_{1}}{2|\psi_{1}|^{2}}
\Big[ \sin(\theta_{3}+\theta_{2}-\theta_{1})-\sin(\theta_{3}-\theta_{2}+\theta_{1})- \sin(\theta_{1}+\theta_{2}-\theta_{3})    \Big]+ \\
\frac{\psi_{1}}{2|\psi_{1}|}\Big[ \cos(\theta_{3}+\theta_{2}-\theta_{1}) + \cos(\theta_{3}-\theta_{2}+\theta_{1})+\cos(\theta_{1}+\theta_{2}-\theta_{3})  \Big]
\end{multline}
replacing the last Eq. in Eq. (\ref{f1}) 

\begin{multline*}
    2\gamma|\psi_{2}||\psi_{3}\Big[\frac{i\psi_{1}}{2|\psi_{1}|}\Big]
\Big[ \sin(\theta_{3}+\theta_{2}-\theta_{1})-\sin(\theta_{3}-\theta_{2}+\theta_{1})- \sin(\theta_{1}+\theta_{2}-\theta_{3})    \Big]+ \\
 2\gamma|\psi_{2}||\psi_{3}\Big[\frac{\psi_{1}}{2|\psi_{1}|}\Big]\Big[ \cos(\theta_{3}+\theta_{2}-\theta_{1}) + \cos(\theta_{3}-\theta_{2}+\theta_{1})+\cos(\theta_{1}+\theta_{2}-\theta_{3})  \Big],
\end{multline*}
\begin{multline*}
    \gamma|\psi_{2}||\psi_{3}\Big[\frac{\psi_{1}}{|\psi_{1}|}\Big]
\Big[i( \sin(\theta_{3}+\theta_{2}-\theta_{1})-\sin(\theta_{3}-\theta_{2}+\theta_{1})- \sin(\theta_{1}+\theta_{2}-\theta_{3}) ) + \\
\cos(\theta_{3}+\theta_{2}-\theta_{1}) + \cos(\theta_{3}-\theta_{2}+\theta_{1})+\cos(\theta_{1}+\theta_{2}-\theta_{3})  \Big]
\end{multline*}
finally, we can define $\hat{\gamma}_{23}$, 
\begin{multline*}
   \hat{\gamma}_{23} = \gamma|\psi_{2}||\psi_{3}|\epsilon^{i\theta_{1}}
\Big[i( \sin(\theta_{3}+\theta_{2}-\theta_{1})-\sin(\theta_{3}-\theta_{2}+\theta_{1})- \sin(\theta_{1}+\theta_{2}-\theta_{3}) ) + \\
\cos(\theta_{3}+\theta_{2}-\theta_{1}) + \cos(\theta_{3}-\theta_{2}+\theta_{1})+\cos(\theta_{1}+\theta_{2}-\theta_{3})  \Big]
\end{multline*}
and 
\begin{empheq}[box=\fbox]{align}
 \hat{\gamma}_{23}  = \gamma|\psi_{2}||\psi_{3}|
\Big[i( \sin(\theta_{3}+\theta_{2})-\sin(\theta_{3}-\theta_{2})- \sin(\theta_{2}-\theta_{3}) ) +
\cos(\theta_{3}+\theta_{2}) + \cos(\theta_{3}-\theta_{2})+\cos(\theta_{2}-\theta_{3})  \Big]
\end{empheq}
\label{Gamma1}

    \item The second functional derivative, $\delta\psi^{*}_{2}=\frac{\partial}{\partial\psi^{*}_{2}}$

{\small\begin{eqnarray}
2\gamma |\psi_{1}||\psi_{3}|\Big[\delta {\psi_{2}^{*}} [|\psi_{2}|\cos(\theta_{3}+\theta_{2}-\theta_{1})]+\delta {\psi_{2}^{*}} [|\psi_{2}|\cos(\theta_{3}-\theta_{2}+\theta_{1})]+\delta {\psi_{2}^{*}} [|\psi_{2}|\cos(\theta_{1}+\theta_{2}-\theta_{3})]   \Big]
\label{f1b}
\end{eqnarray}}
for the firth term of the last Eq. 
\begin{equation*}
\frac{\partial}{\partial \psi_{2}^{*}} \Big[|\psi_{2}|\cos(\theta_{3}+\theta_{2}-\theta_{1})\Big] =   |\psi_{2}|\Big[-\sin( \theta_{3}+\theta_{2}-\theta_{1}) \cdot (\frac{\partial\theta_{2}}{\partial\psi_{2}^{*}}) \Big]+\cos(\theta_{3}+\theta_{2}-\theta_{1})\Big[  \frac{\psi_{2}}{2|\psi_{2}|} \Big],
\end{equation*}
\begin{equation*}
\frac{\partial}{\partial \psi_{2}^{*}} \Big[|\psi_{2}|\cos(\theta_{3}+\theta_{2}-\theta_{1})\Big] =   -|\psi_{2}|\sin( \theta_{3}+\theta_{2}-\theta_{1})  (\frac{\partial\theta_{2}}{\partial\psi_{2}^{*}}) + \frac{\psi_{2}}{2|\psi_{2}|} \cos(\theta_{3}+\theta_{2}-\theta_{1})
\end{equation*}
and with the phases, see the appendix (\ref{apendiceB}), we can write the last Eq. as
\begin{equation}
\frac{\partial}{\partial \psi_{2}^{*}} \Big[|\psi_{2}|\cos(\theta_{3}+\theta_{2}-\theta_{1})\Big] =  -i|\psi_{2}|\frac{\psi_{2}}{2|\psi_{2}|^{2}}  \sin( \theta_{3}+\theta_{2}-\theta_{1})  + \frac{\psi_{2}}{2|\psi_{2}|} \cos(\theta_{3}+\theta_{2}-\theta_{1})
\label{A3b}
\end{equation}
for the second term
\begin{equation*}
    \delta\psi_{2}^{*}[|\psi_{2}|\cos(\theta_{3}-\theta_{2}+\theta_{1})]=|\psi_{2}|\cdot -\sin(\theta_{3}-\theta_{2}+\theta_{1})\left( \frac{-\partial\theta_{2} }{\partial\psi_{2}^{*}}\right)+ \cos( \theta_{3}-\theta_{2}+\theta_{1}) \frac{\psi_{2}}{2|\psi_{2}|},
\end{equation*}
\begin{equation}
    \delta\psi_{2}^{*}[|\psi_{2}|\cos(\theta_{3}-\theta_{2}+\theta_{1})]=i\frac{|\psi_{2}|\psi_{2}}{2|\psi_{2}|^{2}}\cdot \sin(\theta_{3}-\theta_{2}+\theta_{1})+ \frac{\psi_{2}}{2|\psi_{2}|}\cos( \theta_{3}-\theta_{2}+\theta_{1}) 
    \label{A5b}
\end{equation}
and for the third term
\begin{equation*}
    \delta\psi_{2}^{*}[|\psi_{2}|\cos(\theta_{1}+\theta_{2}-\theta_{3})]=|\psi_{2}|\cdot -\sin(\theta_{1}+\theta_{2}-\theta_{3})\left( \frac{\partial\theta_{2} }{\partial\psi_{2}^{*}}\right)+ \cos( \theta_{1}+\theta_{2}-\theta_{1}) \frac{\psi_{2}}{2|\psi_{2}|},
\end{equation*}
\begin{equation}
    \delta\psi_{2}^{*}[|\psi_{2}|\cos(\theta_{1}+\theta_{2}-\theta_{3})]=-i\frac{|\psi_{2}|\psi_{2}}{2|\psi_{2}|^{2}}\cdot \sin(\theta_{1}+\theta_{2}-\theta_{1})+ \frac{\psi_{2}}{2|\psi_{2}|}\cos( \theta_{1}+\theta_{2}-\theta_{3}) 
    \label{A7b}
\end{equation}

Now, grouping the Eqs (\ref{A3b}), (\ref{A5b}) and (\ref{A7b}), we have that

\begin{multline*}
    i|\psi_{2}|\frac{\psi_{2}}{2|\psi_{2}|^{2}}
\Big[ -\sin(\theta_{3}+\theta_{2}-\theta_{1})+\sin(\theta_{3}-\theta_{2}+\theta_{1})- \sin(\theta_{1}+\theta_{2}-\theta_{3})    \Big]+ \\
\frac{\psi_{2}}{2|\psi_{2}|}\Big[ \cos(\theta_{3}+\theta_{2}-\theta_{1}) + \cos(\theta_{3}-\theta_{2}+\theta_{1})+\cos(\theta_{1}+\theta_{2}-\theta_{3})  \Big]
\end{multline*}
replacing the last Eq. in Eq. (\ref{f1b}), 
\begin{multline*}
    2\gamma|\psi_{1}||\psi_{3}|\Big[\frac{i\psi_{2}}{2|\psi_{2}|}\Big]
\Big[ -\sin(\theta_{3}+\theta_{2}-\theta_{1})+\sin(\theta_{3}-\theta_{2}+\theta_{1})- \sin(\theta_{1}+\theta_{2}-\theta_{3})    \Big]+ \\
 2\gamma|\psi_{1}||\psi_{3}\Big[\frac{\psi_{2}}{2|\psi_{2}|}\Big]\Big[ \cos(\theta_{3}+\theta_{2}-\theta_{1}) + \cos(\theta_{3}-\theta_{2}+\theta_{1})+\cos(\theta_{1}+\theta_{2}-\theta_{3})  \Big]
\end{multline*}
\begin{multline*}
    \gamma|\psi_{2}||\psi_{3}\Big[\frac{\psi_{1}}{|\psi_{1}|}\Big]
\Big[i( -\sin(\theta_{3}+\theta_{2}-\theta_{1})+\sin(\theta_{3}-\theta_{2}+\theta_{1})- \sin(\theta_{1}+\theta_{2}-\theta_{3}) ) + \\
\cos(\theta_{3}+\theta_{2}-\theta_{1}) + \cos(\theta_{3}-\theta_{2}+\theta_{1})+\cos(\theta_{1}+\theta_{2}-\theta_{3})  \Big]
\end{multline*}
finally, we can define $\hat{\gamma}_{23}$, 
\begin{multline*}
   \hat{\gamma}_{13} = \gamma|\psi_{1}||\psi_{3}|\epsilon^{i\theta_{2}}
\Big[i( -\sin(\theta_{3}+\theta_{2}-\theta_{1})+\sin(\theta_{3}-\theta_{2}+\theta_{1})-\sin(\theta_{1}+\theta_{2}-\theta_{3}) ) + \\
\cos(\theta_{3}+\theta_{2}-\theta_{1}) + \cos(\theta_{3}-\theta_{2}+\theta_{1})+\cos(\theta_{1}+\theta_{2}-\theta_{3})  \Big]
\end{multline*}

and

\begin{empheq}[box=\fbox]{align}
 \hat{\gamma}_{13} = \gamma|\psi_{1}||\psi_{3}|
\Big[i(\sin(\theta_{3}+\theta_{1})-\sin(\theta_{3}-\theta_{1})-\sin(\theta_{1}-\theta_{3}) ) + 
\cos(\theta_{3}+\theta_{1}) + \cos(\theta_{3}-\theta_{1})+\cos(\theta_{1}-\theta_{3})  \Big]
\end{empheq}
\label{Gamma2}

 \item The third functional, $\delta\psi^{*}_{3}=\frac{\partial}{\partial\psi^{*}_{3}}$
 
 {\small\begin{eqnarray}
2\gamma |\psi_{1}||\psi_{2}|\left[\delta {\psi_{3}^{*}} [|\psi_{3}|\cos(\theta_{3}+\theta_{2}-\theta_{1})]+\delta {\psi_{3}^{*}} [|\psi_{3}|\cos(\theta_{3}-\theta_{2}+\theta_{1})]+\delta {\psi_{3}^{*}} [|\psi_{3}|\cos(\theta_{1}+\theta_{2}-\theta_{3})]      \right]
\label{f1c}
\end{eqnarray}}

for the firth term of the last Eq. 

\begin{equation*}
\frac{\partial}{\partial \psi_{3}^{*}} \Big[|\psi_{3}|\cos(\theta_{3}+\theta_{2}-\theta_{1})\Big] =   |\psi_{3}|\Big[-\sin( \theta_{3}+\theta_{2}-\theta_{1}) \cdot (\frac{\partial\theta_{3}}{\partial\psi_{3}^{*}}) \Big]+\cos(\theta_{3}+\theta_{2}-\theta_{1})\Big[  \frac{\psi_{3}}{2|\psi_{3}|} \Big],
\end{equation*}
\begin{equation*}
\frac{\partial}{\partial \psi_{3}^{*}} \Big[|\psi_{3}|\cos(\theta_{3}+\theta_{2}-\theta_{1})\Big] =   -|\psi_{3}|\sin( \theta_{3}+\theta_{2}-\theta_{1})  (\frac{\partial\theta_{3}}{\partial\psi_{3}^{*}}) + \frac{\psi_{3}}{2|\psi_{3}|} \cos(\theta_{3}+\theta_{2}-\theta_{1})
\end{equation*}
and with the phases, see the appendix (\ref{apendiceB}), we can write the last Eq. as
\begin{equation}
\frac{\partial}{\partial \psi_{3}^{*}} \Big[|\psi_{3}|\cos(\theta_{3}+\theta_{2}-\theta_{1})\Big] =  -i|\psi_{3}|\frac{\psi_{3}}{2|\psi_{3}|^{2}}  \sin( \theta_{3}+\theta_{2}-\theta_{1})  + \frac{\psi_{3}}{2|\psi_{3}|} \cos(\theta_{3}+\theta_{2}-\theta_{1})
\label{A3c}
\end{equation}
for the second term
\begin{equation*}
    \delta\psi_{3}^{*}[|\psi_{3}|\cos(\theta_{3}-\theta_{2}+\theta_{1})]=|\psi_{3}|\cdot -\sin(\theta_{3}-\theta_{2}+\theta_{1})\left( \frac{\partial\theta_{3} }{\partial\psi_{3}^{*}}\right)+ \cos( \theta_{3}-\theta_{2}+\theta_{1}) \frac{\psi_{3}}{2|\psi_{3}|},
\end{equation*}
\begin{equation}
    \delta\psi_{3}^{*}[|\psi_{3}|\cos(\theta_{3}-\theta_{2}+\theta_{1})]=-i\frac{|\psi_{3}|\psi_{3}}{2|\psi_{3}|^{2}}\cdot \sin(\theta_{3}-\theta_{2}+\theta_{1})+ \frac{\psi_{2}}{2|\psi_{2}|}\cos( \theta_{3}-\theta_{2}+\theta_{1}) 
    \label{A5c}
\end{equation}
and for the third term
\begin{equation*}
    \delta\psi_{3}^{*}[|\psi_{3}|\cos(\theta_{1}+\theta_{2}-\theta_{3})]=|\psi_{3}|\cdot -\sin(\theta_{1}+\theta_{2}-\theta_{3})\left( \frac{-\partial\theta_{3} }{\partial\psi_{3}^{*}}\right)+ \cos( \theta_{1}+\theta_{2}-\theta_{1}) \frac{\psi_{3}}{2|\psi_{3}|},
\end{equation*}
\begin{equation}
    \delta\psi_{3}^{*}[|\psi_{3}|\cos(\theta_{1}+\theta_{2}-\theta_{3})]=i\frac{|\psi_{3}|\psi_{3}}{2|\psi_{3}|^{2}}\cdot \sin(\theta_{1}+\theta_{2}-\theta_{1})+ \frac{\psi_{3}}{2|\psi_{3}|}\cos( \theta_{1}+\theta_{2}-\theta_{3}) 
    \label{A7c}
\end{equation}
Now, grouping the Eqs (\ref{A3c}), (\ref{A5c}) and (\ref{A7c}), we have that
\begin{multline*}
    i|\psi_{2}|\frac{\psi_{2}}{2|\psi_{2}|^{2}}
\Big[ -\sin(\theta_{3}+\theta_{2}-\theta_{1})+\sin(\theta_{3}-\theta_{2}+\theta_{1})- \sin(\theta_{1}+\theta_{2}-\theta_{3})    \Big]+ \\
\frac{\psi_{2}}{2|\psi_{2}|}\Big[ \cos(\theta_{3}+\theta_{2}-\theta_{1}) + \cos(\theta_{3}-\theta_{2}+\theta_{1})+\cos(\theta_{1}+\theta_{2}-\theta_{3})  \Big]
\end{multline*}
replacing the last Eq. in Eq. (\ref{f1c}), we getting  
\begin{multline*}
    2\gamma|\psi_{1}||\psi_{2}|\Big[\frac{i\psi_{3}}{2|\psi_{3}|}\Big]
\Big[ -\sin(\theta_{3}+\theta_{2}-\theta_{1})+\sin(\theta_{3}-\theta_{2}+\theta_{1})- \sin(\theta_{1}+\theta_{2}-\theta_{3})    \Big]+ \\
 2\gamma|\psi_{1}||\psi_{2}\Big[\frac{\psi_{3}}{2|\psi_{3}|}\Big]\Big[ \cos(\theta_{3}+\theta_{2}-\theta_{1}) + \cos(\theta_{3}-\theta_{2}+\theta_{1})+\cos(\theta_{1}+\theta_{2}-\theta_{3})  \Big],
\end{multline*}
\begin{multline*}
    \gamma|\psi_{1}||\psi_{2}\Big[\frac{\psi_{3}}{|\psi_{3}|}\Big]
\Big[i( -\sin(\theta_{3}+\theta_{2}-\theta_{1})+\sin(\theta_{3}-\theta_{2}+\theta_{1})- \sin(\theta_{1}+\theta_{2}-\theta_{3}) ) + \\
\cos(\theta_{3}+\theta_{2}-\theta_{1}) + \cos(\theta_{3}-\theta_{2}+\theta_{1})+\cos(\theta_{1}+\theta_{2}-\theta_{3})  \Big]
\end{multline*}
finally, we can define $\hat{\gamma}_{23}$, 
\begin{multline}
   \hat{\gamma}_{12} = \gamma|\psi_{1}||\psi_{2}|\epsilon^{i\theta_{3}}
\Big[i( -\sin(\theta_{3}+\theta_{2}-\theta_{1})+\sin(\theta_{3}-\theta_{2}+\theta_{1})-\sin(\theta_{1}+\theta_{2}-\theta_{3}) ) + \\
\cos(\theta_{3}+\theta_{2}-\theta_{1}) + \cos(\theta_{3}-\theta_{2}+\theta_{1})+\cos(\theta_{1}+\theta_{2}-\theta_{3})  \Big],
\end{multline}
 
\begin{empheq}[box=\fbox]{align}
 \hat{\gamma}_{12} = \gamma|\psi_{1}||\psi_{2}|
\Big[i(\theta_{2}+\theta_{1})-\sin(\theta_{1}-\theta_{2})-\sin(\theta_{2}-\theta_{1}) ) + 
\cos(\theta_{2}+\theta_{1}) + \cos(\theta_{2}-\theta_{1})+\cos(\theta_{1}-\theta_{2})  \Big].
\end{empheq}
\label{Gamma3}




\section{Phases}\label{apendiceB}
The complex order parameter for $i$-band is given by 
\begin{equation*}
    \psi_{i}^{*}=|\psi_{i}|e^{-i\theta_{i}}.
\end{equation*}
This order parameter can be represented as
\begin{equation*}
    \ln({\psi_{i}^{*}})= \ln{ [|\psi_{i}|\epsilon^{-i\theta_{i}}]},
\end{equation*}
\begin{equation*}
    \ln({\psi_{i}^{*}})= \ln{ [|\psi_{i}|+\ln[\epsilon^{-i\theta_{i}}]}
\end{equation*}
\begin{equation*}
    \ln({\psi_{i}^{*}})= \ln{ [|\psi_{i}|}-i\theta_{i}
\end{equation*}
them, the $\theta_{i}$ is 
\begin{equation*}
    \theta_{i}=-i\ln{ [|\psi_{1}|}+i\ln{\psi_{1}^{*}}
\end{equation*}

Applying the functional derivate to $\theta_{i}$, we have 

\begin{empheq}[box=\fbox]{align}
    \frac{\partial\theta_{i}}{\partial_{\psi_{i}^{*}}}=i\left( \frac{1}{|\psi_{i}|} \Big[\frac{\psi_{1}}{2|\psi_{i}|}\Big] \right)=i\frac{\psi_{1}}{2|\psi_{i}|^{2}}.
\end{empheq}

\end{itemize}
\end{widetext}


\end{document}